\documentclass[twocolumn]{article}

\usepackage{graphicx} % For including images
\usepackage{amsmath} % For math formulas
\usepackage{lipsum} % For generating dummy text
\usepackage{authblk} % For handling multiple authors and affiliations

\usepackage{amssymb} % Provides access to more math symbols and fonts

\usepackage{multirow} % for merging rows
\usepackage{dcolumn}% Align table columns on decimal point
\usepackage{bm} % For bold math symbols
\usepackage[utf8]{inputenc} % For UTF-8 encoding, allowing special characters
\usepackage[T1]{fontenc} % For better font encoding, improving character display
\usepackage{mathptmx} % For using Times New Roman math fonts
\usepackage{etoolbox} % Provides tools for conditional execution and patching
\usepackage{xcolor} % Provides tools for conditional execution and patching

\usepackage{tabularray} % Enhanced table creation with flexible options
\UseTblrLibrary{diagbox}  % A library for the tabularray package to create diagonal lines in table cells
\usepackage[colorlinks,citecolor=black]{hyperref} % For creating hyperlinks in the document
\usepackage[backend=biber, style=numeric-comp, sorting=none, doi=false, url=false, eprint=false]{biblatex} % For managing the bibliography with Biber
\AtBeginEnvironment{tabular}{\small} % % Define the default font size for tables; Use \small, \footnotesize, \scriptsize, etc.

\usepackage{xr} % for referencing objects in different file
\usepackage{geometry} % for margines
\title{Active learning and explicit electrostatics enable accurate modeling of electrolytes}

\author[1]{Olga Chalykh}
\author[1,2]{Mikhail Polovinkin}
\author[1,2]{Dmitry Korogod}
\author[1,2,3]{Nikita Rybin}
\author[1,2,*]{Alexander Shapeev}%\thanks{Corresponding author: a.shapeev@skoltech.ru}}

\affil[1]{Skolkovo Institute of Science and Technology, Bolshoy boulevard 30, Moscow, 143026, Russia}
\affil[2]{Digital Materials LLC, Odintsovo, Kutuzovskaya str. 4A, Odintsovo, 143001, Russia}
\affil[3]{Moscow Engineering Physics Institute, Kashirskoe Highway 31, Moscow, 115409, Russia}
\affil[*]{Corresponding author: Alexander Shapeev, a.shapeev@skoltech.ru}

\date{\today}

\begin{document}

\maketitle

\begin{abstract}

Machine learning interatomic potentials (MLIPs) combine near-\textit{ab initio} accuracy with the computational efficiency of classical force fields, making them promising tools for modeling molecular liquids and electrolytes. Their predictive accuracy depends strongly on the diversity of the training set, while the explicit treatment of electrostatics is increasingly recognized as an important factor in improving their performance. In this work, we assess the impact of incorporating explicit electrostatics and environment-dependence of atomic charges by comparing short-ranged Moment Tensor Potentials (MTPs) with electrostatics-augmented MTPs, including fixed-charge and environment-dependent charge models. To this end, we employ D-optimality-based active learning (AL) to train short-ranged MTPs and extend the AL procedure, previously unavailable for long-range electrostatics-augmented MTPs, to these models, enabling largely automated generation of diverse training sets for pure ethylene carbonate (EC) and ethyl methyl carbonate (EMC), their mixtures, and LiPF$_6$ solutions. For EC, EMC, and their mixtures, electrostatics-augmented MTPs achieve accuracy comparable to or better than that of their short-ranged counterparts while requiring fewer machine-learned parameters. For LiPF$_6$ solutions, however, a notable improvement in accuracy is achieved only with environment-dependent charges, which also enhance stability in large-scale molecular dynamics simulations, unlike the fixed-charge model. The resulting short-ranged MTPs and environment-dependent-charge models reproduce experimental densities with mean absolute deviations of 0.2–6\% and ionic conductivities with deviations of 11–15\%, while capturing the experimentally observed temperature dependence of conductivity.

\end{abstract}

\section{Introduction}

Performance of lithium-ion batteries is strongly influenced by the electrolyte, which facilitates lithium-ion transport between electrodes.
Optimizing the ionic conductivity of the electrolyte requires solvents that combine high dielectric permittivity to enhance salt dissociation and low viscosity to facilitate lithium-ion migration~\cite{franco2019boosting}.
These requirements impose strong constraints, leading to the use of mixtures of multiple organic solvents to harness complementary physicochemical properties~\cite{de2024calisol}.
Thus, optimizing the electrolyte composition requires extensive screening of salt--solvent combinations.
This makes computational modeling valuable for streamlining experimental screening by identifying electrolyte candidates that show promise in simulation results~\cite{franco2019boosting}. 

Molecular dynamics (MD) simulations are a primary tool for electrolyte modeling, with their predictive power determined by the interatomic interaction model. \textit{Ab initio} MD (AIMD) can, in principle, accurately account for all interactions. This makes AIMD a valuable tool for studying the structural properties of the Li$^+$ solvation shell that strongly influences ion transport. However, AIMD simulations are limited to a few hundred atoms and timescales of only several tens of picoseconds~\cite{pham2019ab}. These limitations become particularly significant when evaluating transport properties~\cite{ong2015lithium}.

Classical force fields (FFs) enable MD simulations at significantly larger time and size scales.
FFs enable the investigation of the structure and dynamics of the Li$^+$ solvation shell, mobility of ions, and the evaluation of the transport properties~\cite{franco2019boosting,Hou2021}. However, ionic conductivity cannot be accurately predicted by simple nonpolarizable FFs as they show deviations by factors of 2--10~\cite{franco2019boosting,Hou2021}. Including polarization effects in FFs was demonstrated to leverage their accuracy, as polarized FFs yield ionic conductivity with a maximum error of only a few dozen percent~\cite{borodin2006development,borodin2009quantum}. However, polarizable FFs still lack accuracy in describing the solvation shell structure~\cite{ong2015lithium}. 

Machine learning interatomic potentials (MLIPs) offer a promising path to retain AIMD accuracy at a computational cost comparable to FFs. 
In particular, the emergence of universal MLIPs capable of modeling diverse electrolytes is poised to transform electrolyte screening efficiency, as demonstrated in recent studies~\cite{batatia2023foundation,ju2025application,dajnowicz2022high,willow2024sparse}, although fine-tuning remains necessary~\cite{ju2025application}.
An alternative approach is the development of system-specific MLIPs for pure solvents and electrolytes~\cite{magduau2023machine,niblett2025transferability}. These models can rapidly achieve high accuracy for target systems, making them valuable for systems underrepresented in universal MLIP training sets and for testing new model architectures. Moreover, their transferability across solvent compositions and temperatures~\cite{magduau2023machine,niblett2025transferability} enables electrolyte composition screening, even though system-specific MLIPs generally cover a narrower range of compositions and thermodynamic conditions than universal MLIPs.

However, both system-specific and universal MLIPs face two common challenges. The first challenge is the treatment of long-range interactions, which arises from the inherent locality of MLIPs: interactions are explicitly described only within a finite cutoff radius, while the interactions beyond this range are neglected.
This limitation is particularly relevant for electrolytes, which contain charged species and exhibit strong polarization effects, making long-range electrostatic interactions potentially important for accurately describing their structure and properties.
Several strategies have been developed to address this limitation, including message passing~\cite{Batatia2022mace,Batatia2022Design,kong2026scalable}, the inclusion of global information in MLIP descriptors~\cite{kabylda2023efficient}, and explicit incorporation of electrostatic interactions~\cite{behlercsany2021machine,zhang2022deep,unke2019physnet,yao2018tensormol,ko2023accurate}

At the same time, local MLIPs have also been successfully applied to the modeling of pure solvents and electrolytes~\cite{batatia2023foundation,ju2025application,magduau2023machine,niblett2025transferability}, as the major portion of electrostatic interactions can still be captured within a typical cutoff radius (5\textendash6\AA)~\cite{magduau2023machine}. 
However, it was also shown that for modeling electrolyte solutions, MLIPs with explicit long-range interactions outperform local ones not only in force and energy errors but also by yielding a more qualitatively correct solvation environment for ions~\cite{kocer2024machine}.
Thus, quantifying the impact of the explicit incorporation of electrostatic interactions in modeling polar solvents and liquid electrolytes would help to clarify whether their inclusion is necessary.

Another challenge lies in efficiently generating a dataset for training a system-specific MLIP or fine-tuning a universal MLIP, as it crucially affects the performance of a trained model.
The difficulty of this task is illustrated by a local MLIP developed for the ethylene carbonate (EC)/ethyl methyl carbonate (EMC) binary solvent~\cite{magduau2023machine}. Despite being trained on a diverse bulk dataset spanning a range of densities and temperatures, the model exhibited unphysical density dynamics and failed to maintain a stable liquid phase. Stability was achieved only after iterative manual augmentation of the training set.
A similar procedure was used to construct a training set for fine-tuning a universal MLIP~\cite{ju2025application}. 
Although such manual augmentation may not be necessary for some MLIPs~\cite{niblett2025transferability}, it remains a significant challenge for others.
To automate the selection of structures for inclusion in training sets, various active learning (AL) approaches have been developed. These include methods based on query-by-committee~\cite{smith2018less}, optimization of the entropy of atomic descriptors~\cite{Zapiain2022}, Bayesian uncertainty estimates~\cite{kim2026machine}, and the D-optimality criterion and MaxVol algorithm~\cite{podryabinkin2017active,gubaev2019accelerating}.

In this study, we address two questions: the first addresses the impact of explicitly including electrostatic interactions and the environment dependence of charges on the accuracy of MLIPs in modeling the EC/EMC binary solvent and LiPF$_6$ solution in EC/EMC, and the second concerns an efficient AL-based workflow for generating training sets for these MLIPs.

To address the first question, we compare MLIPs with and without explicit electrostatics while keeping the local MLIP component consistent, thereby isolating the effect of incorporating electrostatics.
These setups can be achieved with a variety of models, including HDNNPs~\cite{behler2021four}, DP models~\cite{zhang2022deep}, MTP models~\cite{Shapeev2016}, and others~\cite{anstine2023machine,behler2021four,batatia2026mace}. Here, we compare the Moment Tensor Potential~\cite{Shapeev2016} (MTP) with versions augmented by an explicit Coulomb interaction term parameterized by machine-learned charges. Specifically, we consider two electrostatics-augmented variants: MTP-QRd~\cite{korogod2026incorporating}, which uses charges fixed for each atomic type, and MTP-EDQRd~\cite{korogod2026edq}, which incorporates environment-dependent charges. This choice allows us to investigate both the impact of explicitly incorporating electrostatics and the additional impact of making the charges environment-dependent.

In this setup, MTP-QRd serves as a simple model for assessing the impact of explicit electrostatics, whereas MTP-EDQRd enables evaluation of the role of environment-dependent charges. Although MTP is a well-established approach that has proven efficient in diverse applications~\cite{zuo2020performance,gubaev2019accelerating,podryabinkin2019accelerating,rybin2025accelerating,chalykh2025moment,rybin2024thermophysical,klimanova2025accelerating,novikov2024towards}, the MTP-QRd and MTP-EDQRd models are newly developed and have only been applied to molecular dimers~\cite{korogod2026incorporating} and, in the case of MTP-EDQRd, additionally to NaCl and PbTiO$_3$~\cite{korogod2026edq}.
We benchmark the MTP, MTP-QRd, and MTP-EDQRd potentials on EC/EMC mixtures and LiPF$_6$ solution in EC/EMC, evaluating their accuracy in atomic forces, densities, radial distribution functions, and ionic conductivities against both \textit{ab initio} calculations and experimental data.

To address the second question, we extend D-optimality-based AL~\cite{podryabinkin2017active,goreinov2010-maxvol,gubaev2019accelerating,Novikov2021}, originally developed for short-ranged MTPs, to long-range electrostatics-augmented MTPs, for which AL was previously unavailable. We further incorporate AL into a training workflow to automate training sample selection.
We first validate whether our AL method generates training sets that yield MTPs capable of maintaining a stable liquid density in MD simulations. Second, we incorporate AL into a pipeline for constructing MLIPs transferable across EC/EMC compositions, enabling us to sample the compositional space between pure components and obtain a compact, representative training set.
Finally, we apply an AL-based pipeline to generate a training set for a 1M~LiPF$_6$ solution in an 3EC:7EMC (molar ratio) binary solvent. This system poses a greater challenge since the salt introduces three additional chemical elements, significantly increasing the diversity of local environments. Moreover, the solvation environments of the Li$^+$ ions exhibit lifetimes on the order of hundreds of picoseconds~\cite{Hou2021} and are highly variable. These factors make comprehensive sampling of the configurational space challenging on the timescales of MD simulations employed in AL. This limitation is particularly critical for ionic conductivity calculations, which require long nanosecond-scale simulations that may drive the system beyond the training domain.

\section{Methods}

\subsection{Charge Redistribution}\label{methods_QRd}
Charge Redistribution (QRd) is a long-range model employed for explicit electrostatic interaction treatment~\cite{korogod2026incorporating}. The QRd energy corresponds to the interaction of point charges centered on atoms:

\begin{equation}
    \label{eq:qrd_en}
    E^{\rm QRd} = \sum_{i<j}\frac{q_iq_j}{\vert r_{ij}\vert},
\end{equation}
where $q_i,q_j$ are the charges of $i$-th and $j$-th atoms. These charges are predicted with the QRd model, using the following expression:

\begin{equation}
q_i(\bm{z,a}) = q_i(\bm{z},(\bm{b,s})) = b_{z_i} + s_{z_i}\frac{Q_{\rm total} - \sum_jb_{z_j}}{\sum_j s_{z_j}},
\end{equation}
where $\bm{b}$ and $\bm{s}$ are vectors of the model parameters and $Q_{\rm total}$ is the total charge of the system, allowing the conservation of the total charge.

This scheme has an intrinsic limitation as the predicted charges do not depend on the atomic environment and are solely dependent on the chemical composition of the system (atomic types of all atoms comprising the system). Specifically, this leads to the same charges for chemically different atoms of the same type (for example, carbonyl oxygen and ether oxygen). Thus, predicted long-range interaction would be biased by these inaccuracies.

The QRd term can be applied on the top of a short-range MLIP, resulting in MTP-QRd potential with explicit long-range electrostatic interactions. The corresponding total energy expression for MTP-QRd model thus can be represented as:

\begin{equation}
    \label{eq:total_e_with_qrd}
    E^{\rm MTP-QRd}\left(\bm x, \left(\bm \theta, \bm a\right)\right) = E^{\rm MTP}\left(\bm x,\bm \theta\right) + E^{\rm QRd}(\bm x,\bm a),
\end{equation}
where $\bm \theta$ and $\bm a$ are the MTP and QRd parameters (see Supplementary Information for details on MTP architecture), and $\bm x$ is an atomic configuration.

\subsection{Environment-Dependent Charge Redistribution}\label{Methodology_edqrd}

Incorporating environment-dependent charges overcomes the critical QRd model limitation---the inability to distinguish chemically different atoms of the same type (e.g., ether and carbonyl oxygens). In this scheme, the type-dependent parameter $b$ present in the QRd scheme is replaced by an environment-dependent term $V$, resulting in predicting charges based on both chemical composition and local atomic environment~\cite{korogod2026edq}. Specifically, charges are expressed as:

\begin{equation}
    \label{eq:edqrd_charges}
    \begin{split}
    q_i(\bm x, \bm a) = q_i(\bm x, (\bm p, \bm s))\\\\
     = V(\bm n_i, \bm p) + s_{z_i} \frac{Q_{\rm total} - \sum_{j=1}^N V(\bm n_j, \bm p)}{\sum_{j=1}^N s_{z_j}},        
    \end{split}
\end{equation}

where $\bm x$ is the configuration, $\bm p$ represents short-ranged model parameters, $\bm s$ is a vector of machine-learned parameters associated with charge redistribution, and $n_i$ is the local atomic neighborhood of the atom whose charge is being predicted.

In this study, we extend the treatment of charges by using in EDQRd Gaussian-broadened charges instead of the point charges used in the QRd implementation. The EDQRd energy term can thus be written as follows:

\begin{equation}
    \label{eq:edqrd_en}
    E^{\rm EDQRd} = \sum_{i<j}\frac{q_iq_j}{\vert r_{ij}\vert}\operatorname{erf}{\left(\frac{\vert r_{ij}\vert}{\sqrt{2\left(d_{z_i}^2+d_{z_j}^2\right)}}\right)},
\end{equation}

where $d_{z_i}$ is a gaussian width of atom of type $z_i$ and $\operatorname{erf}$ is the error function. In this work, we used the same value of $0.5$~\AA \;as a gaussian width for all atomic types. We also note that sums in both~\eqref{eq:qrd_en} and~\eqref{eq:edqrd_en} were computed using Ewald summation~\cite{ClassicalEwaldSummation}.

Similar to QRd, EDQRd can be combined with a short-ranged model, with the total energy expression constructed analogously to Eq.~\ref{eq:total_e_with_qrd}. In addition to incorporating the environment dependence of charges, the complexity of the short-ranged model predicting term $V$ in Eq.~\ref{eq:edqrd_charges} is tunable to match the system complexity. In this work, we employed level-12 MTP as the short-ranged model for this purpose.

We note that while the EDQRd represents a significant advancement over the QRd model, it is not sufficient for modeling systems with charges dependent on the structure outside the short-ranged model's cutoff (e.g., for modeling protonation of organic molecules with a large $\pi$-conjugated system). In this case, global structural information should be taken into account as was implemented in 4-generation High-Dimensional Neural Network Potentials~\cite{behler2021four,ko2021fourth}.

\subsection{Fitting}
\label{Methodology_fitting}

To find the optimal parameters of the MLIP $\bm \Omega$ ($\bm \Omega = \bm \theta$ for MTP and $\bm \Omega = \left(\bm \theta, \bm a\right)$ for MTP-QRd and MTP-EDQRd) we minimize the loss function with respect to the parameters of the model:

\begin{equation}
\begin{array}{c}
     \displaystyle
    \mathcal{L}\left(\bm \Omega \right) = \sum_{k=1}^{K} \Big[ \frac{w_e}{N_k} \left(E^{\rm {MLIP}}\left(\bm x_k,\bm \Omega\right) - E^{\rm DFT}(\bm x_k)\right)^2 + 
\\
\displaystyle    
w_f\sum_{i}^{N}\sum_{l=1}^3 \left( F_{i,l}^{\rm {MLIP}}\left(\bm x_k, \bm \Omega\right) - F_{i,l}^{\rm DFT}(\bm x_k)\right)^2 \Big], 
\end{array}
\end{equation}
where $E^{\rm MLIP}$ and $F^{\rm MLIP}$ are MLIP-predicted energies and forces, $\bm{x_k}$ is $k$-th configuration of training set, possessing energies $E^{\rm DFT}$ and forces $F_{i,l}^{\rm DFT}$ acting on $i$-th atom ($l=1,2,3$ is the component of force), $N_k$ is the number of atoms in $\bm{x_k}$, and $w_e$, $w_f$ are non-negative weights of the loss function terms, set to $1 \;\text{eV}^{-2}$ and $0.01 \; \left(\text{eV}/\text{\AA}\right)^{-2}$ respectively.

Prior to loss function minimization, parameters of the potential are randomly sampled from the uniform distribution within the range -1 to 1. Minimization was performed using Broyden-Fletcher-Goldfarb-Shanno (BFGS) algorithm.

We also note that, in this work, the parameters of the QRd and EDQRd terms are optimized to minimize energy and force errors. Consequently, the charges predicted by QRd and EDQRd should be treated primarily as latent variables rather than as physically meaningful quantities for direct use. However, both frameworks also support fitting against reference atomic charges. Adopting this scheme would yield charge predictions consistent with a chosen charge partitioning method, which may not correspond to those optimal for minimal energy and force errors. This approach was not adopted in the present work due to an emphasis on primarily energy- and force-related properties.

\subsection{Active Learning}\label{Methods_AL}
 
After an MLIP is trained and the vector of the optimal parameters ${\bm \Omega} = \left( \Omega_1, \ldots, \Omega_m\right)$ is found, we can construct the matrix of derivatives of energies of a set of $K$ configurations $\bm{x}$ in the training set by training parameters:

\[
\mathsf{B}=\left(\begin{matrix}
\frac{\partial E^{\rm {MLIP} }_1}{\partial \Omega_1}({\bm{x_1}, \bm {\Omega}}) & \ldots & \frac{\partial E^{\rm {MLIP}}_1}{\partial \Omega_m}({\bm{x_1}, \bm {\Omega}}) \\
\vdots & & \vdots \\
\frac{\partial E^{\rm {MLIP}}_K}{\partial \Omega_1}({\bm{x_K}, \bm {\Omega}}) & \ldots & \frac{\partial E^{\rm {MLIP}}_K}{\partial \Omega_m}({\bm{x_K}, \bm {\Omega}}) 
\end{matrix}\right).
\]

Now we are interested in finding submatrix $A$ of matrix $B$ with the maximum absolute value of the determinant (maximum volume). For this, we select a set of the most linearly independent rows of $B$ using MaxVol algorithm~\cite{goreinov2010-maxvol}.

The matrix A is then used to calculate extrapolation grade $\gamma$ for each new configuration ${\bm x^*}$  encountered in MD during AL:
\begin{equation}
\label{eq:extrapol_grade}
    \gamma({\bm x^*}) = \max\limits_{1 \le j \le m} |c_j|,
\end{equation}
where the vector $\textbf{c} = (c_1, \ldots, c_m)$ for the configuration ${\bm x^*}$ is computed as

\begin{equation}
\label{eq:c}
    \textbf{c}^\top = \Bigl(\frac{\partial E^{\rm {MLIP}}}{\partial \Omega_1} ({\bm x}^*, \bm {\Omega}), \dots, \frac{\partial E^{\rm {MLIP}}}{\partial \Omega_m} ({\bm x}^*, \bm {\Omega}) \Bigr) A^{-1}.
\end{equation}

Here, two thresholds are introduced: $\gamma_{\rm save} \approx 2$ and $\gamma_{\rm break} \approx 10$, thus defining range  $\gamma_{\rm save}<\gamma<\gamma_{\rm break}$,  where corresponding configurations are sampled and a threshold $\gamma_{\rm break}$ as a stop condition for MD. The overall AL algorithm can be outlined as follows:

\begin{enumerate}
    \item  MD simulation of predefined length, from which new configurations are sampled if their $\gamma$ falls into range from $\gamma_{\rm save}$ to  $\gamma_{\rm break}$, and while there is no configuration with $\gamma > \gamma_{\rm{break}}$ encountered.

    \item Selection among the saved configurations of those that maximize the volume of the matrix $A$ to be added to the training set.
    
    \item Single point \textit{ab initio} calculation of selected configurations.
    
    \item  MLIP retraining and update of the matrix $A$.
\end{enumerate}

These steps are repeated until no configuration is sampled during MD. For all MD runs during AL, trajectories length was set to 100~ps, with 1~fs timestep and conducted in NPT ensemble at 300~K and 1~atm conditions.

We additionally note that the extrapolation grade has been shown to correlate with MLIP prediction errors~\cite{podryabinkin2017active,novikov2020mlip}, providing an upper bound on the expected errors~\cite{lysogorskiy2023active} and thus serving as a reliable criterion for selecting new structures in AL. Furthermore, it can be used to assess AL convergence by analyzing the number of newly selected structures and the MD simulation time before a configuration with a grade exceeding $\gamma_{\rm break}$ is encountered~\cite{novikov2020mlip,lysogorskiy2023active}. Additionally, the extrapolation grade can be used during MD simulations to detect the occurrence of rare structures, as described in Ref.~\cite{novoselov2019moment} for vacancy diffusion in metals.

\subsection{\textit{Ab initio} calculations}

All \textit{ab initio} calculations were performed using Vienna Ab initio Simulation Package (VASP)~\cite{kresse1996software} with density functional theory (DFT), Perdew-Burke-Ernzerhof~\cite{perdew1996generalized} exchange-correlation functional, and the D3 dispersion correction with zero-damping function~\cite{grimme2010consistent}. Previous studies have identified this level of theory as providing an optimal balance between accuracy and computational efficiency for the EC/EMC binary solvent~\cite{magduau2023machine}; PBE has also been widely used for AIMD simulations of liquid electrolytes, with dispersion corrections recommended to further improve accuracy~\cite{haskins2015ab,he2019density,gong2026theoretical}.
All calculations were performed at the $\Gamma$-point only to ensure robustness due to a large number of calculations to be performed. We used the projector-augmented-wave (PAW) method, with 1, 3, 4, 6, 7, and 5 valence electrons for H, Li, C, O, F, and P respectively. The tolerance for the electronic density convergence during the self-consistent field loop was set at 10$^{-6}$ eV. For the simulations of pure EC and EMC and their mixtures we used a plane wave kinetic energy cutoff of 550~eV, while for the simulations of LiPF$_6$ solution in EC/EMC, we applied a higher cutoff of 750~eV, with the cutoffs in both cases being 150 and 250~eV higher, respectively, than the minimum required by the pseudopotentials used.

\section{Results and discussion}

\subsection{Modeling of pure EC and EMC}

\begin{figure*}[!ht]
    \centering
    \includegraphics[width=1\linewidth]{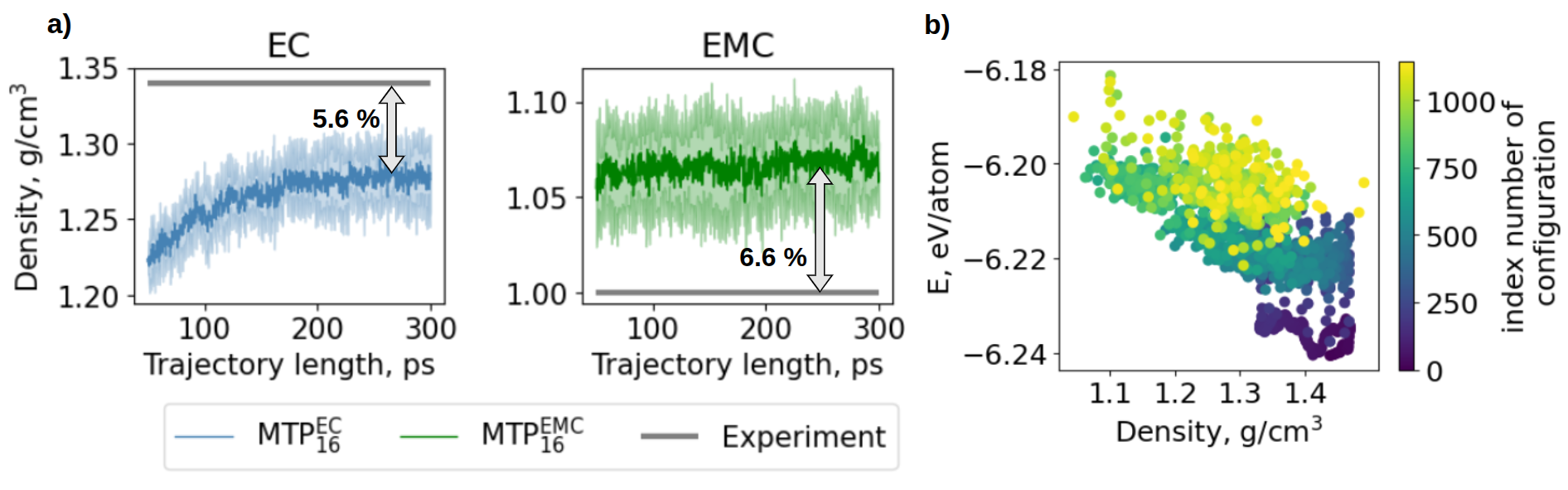}
    \caption{a) Mean densities predicted by an ensemble of 3 MTPs, together with standard deviation due to random initialization of the MLIP parameters, compared to literature values~\cite{magduau2023machine} obtained by interpolating experimental data~\cite{johnson1985properties,wang2022current}; b) energy-density diagram for the training set of pure EC molecular liquid. The color coding represents index number $N$ of the sample in the training set. First 200 samples were extracted from AIMD, after that samples were selected in the active learning loop.}
    \label{fig:pure_EC_EMC}
\end{figure*}

Since the feasibility of subsequent simulations depends on the robustness of our AL-driven training approach, we first assessed whether the employed AL (see Section \ref{Methods_AL}) can yield MLIPs that produce stable and accurate densities in NPT simulations. We began this assessment with pure EC and EMC systems at 300~K, where EC is in a metastable liquid state and EMC is in the liquid phase.
Prior to AL, we constructed an initial training set from AIMD simulations carried out in the NVT ensemble at 400~K for 2~ps with a 1~fs timestep. We uniformly sampled 200 configurations from these simulations and used them to train an initial MLIP, which served as the starting point for AL.
AL-driven training was then performed during ten parallel MD trajectories initialized from configurations sampled from the initial training set.
This pipeline was employed for MTPs with a cutoff of 5~\AA, as interactions beyond this distance are comparable to fitting errors and increasing the cutoff would raise the computational cost without improving accuracy~\cite{magduau2023machine} (this choice was also validated for LiPF$_6$ EC/EMC; see Supplementary Information).
 
For pure EC and EMC, we actively trained MTPs of level 16 (MTP$_{16}^{\rm EC}$ and MTP$_{16}^{\rm EMC}$) (see Supplementary Information, Section \ref{methods_mtp} for a definition of the MTP level) and obtained an ensemble of three potentials for each system to assess uncertainty arising from random initialization of potential parameters (see Section \ref{Methodology_fitting}).
The AL-generated training sets contained between 900 and 1400 configurations, each consisting of 16 EC molecules or 8 EMC molecules, which was sufficient to obtain MTPs producing stable densities in MD simulations without unphysical density fluctuations (Fig.~\ref{fig:pure_EC_EMC}~(a)). Specifically, the obtained deviation from experiment was about 6\% at 300~K on average in the MLIP ensemble, and the associated uncertainty due to random initialization of the MLIP parameters was approximately 2\%, demonstrating the reproducibility of the MLIP accuracy and the robustness of both the AL procedure and the trained potentials.

Furthermore, the training sets collected during AL cover a wide range of densities, analogous to training sets obtained through iterative protocols with manual augmentation using isotropically inflated and deflated structures~\cite{magduau2023machine}.
Fig.~\ref{fig:pure_EC_EMC}~(b) shows the energy-density distribution of configurations from the training set for one of the MTP$_{16}^{\rm EC}$ models. The first 200 samples, obtained from AIMD, occupy the low-energy, high-density region, consistent with sampling near energy minima.
As the configuration index increases, the density range broadens; however, the system remains in the liquid phase, indicating that AL-MD remains within a physically relevant region of configurational space.
This behavior can be attributed both to the ability of the employed MLIP uncertainty measure in AL to terminate MD simulations before unphysical behavior occurs and to the intrinsic ability of MTP to accurately capture intermolecular interactions, thereby preserving the liquid state without density collapse.

Having established the robustness of AL for pure EC and EMC, we next applied it to construct compositionally transferable potentials for the EC/EMC binary solvent. Using the resulting training sets, we then quantified the impact of explicitly including electrostatic interactions and environment-dependent charges on MLIP accuracy.

\subsection{Modeling of EC/EMC mixtures}

Prior to applying AL to obtain a compositionally transferable MLIP, we constructed an initial training set using the AL-generated datasets for pure EC and EMC. We combined the two datasets and uniformly sampled approximately 20 configurations to train a preliminary MTP. This model was then used with the MaxVol algorithm (see Section \ref{Methods_AL}) to select additional representative configurations from the combined datasets. Together, the uniformly sampled and MaxVol-selected configurations formed an initial training set spanning a broad range of atomic environments, which was used to train the MTP that initialized the AL procedure.
AL was subsequently performed on 11 parallel MD trajectories, starting from 8--16-molecule EC/EMC mixture configurations spanning the full compositional range from pure EC to pure EMC. Although the training set could have been constructed from scratch using EC/EMC mixture configurations, reusing the pure-component datasets reduced the computational cost of training.

The described pipeline was applied to construct ensembles of three level-16 and level-20 MTPs (MTP$_{16}$ and MTP$_{20}$), anticipating that compositionally transferable potentials might require larger models than those sufficient for pure components. 
The actively collected training sets contained approximately 500 configurations for both MTP levels, about two to three times fewer than for pure-component systems, and were dominated by pure EC and EMC structures, with mixture configurations accounting for only about 10\% of the total (see Supplementary Information for details). This reduction likely reflects the broader initial training set and the accuracy of the initial MTP, which kept AL-MD trajectories within a physically relevant, well-sampled region of configurational space, thereby reducing the need to collect additional configurations during AL.

After obtaining the training sets, we proceeded to quantify the impact of explicitly incorporating electrostatic interactions and environment-dependent charges on the accuracy of MTPs. To this end, we employed MTP-QRd, which incorporates electrostatics using fixed charges (see details in Section~\ref{methods_QRd}), and MTP-EDQRd, which employs environment-dependent charges (see details in Section~\ref{Methodology_edqrd}). MTP-QRd therefore probes the effect of introducing electrostatics in the simplest form, whereas MTP-EDQRd further allows assessment of the impact of treating charges as environment-dependent quantities.
For both MTP-QRd and MTP-EDQRd, we used AL to train three potentials of each type with level-16 and level-20 MTP components, yielding four model variants: MTP$_{16}$-QRd, MTP$_{20}$-QRd, MTP$_{16}$-EDQRd, and MTP$_{20}$-EDQRd. AL was initialized using training sets previously collected with standard MTP potentials (details of the model naming convention, MTP levels, numbers of fitting parameters, and other model specifications are provided in the Supplementary Information).

\begin{figure*}[!ht]
    \centering
    \includegraphics[width=1\linewidth]{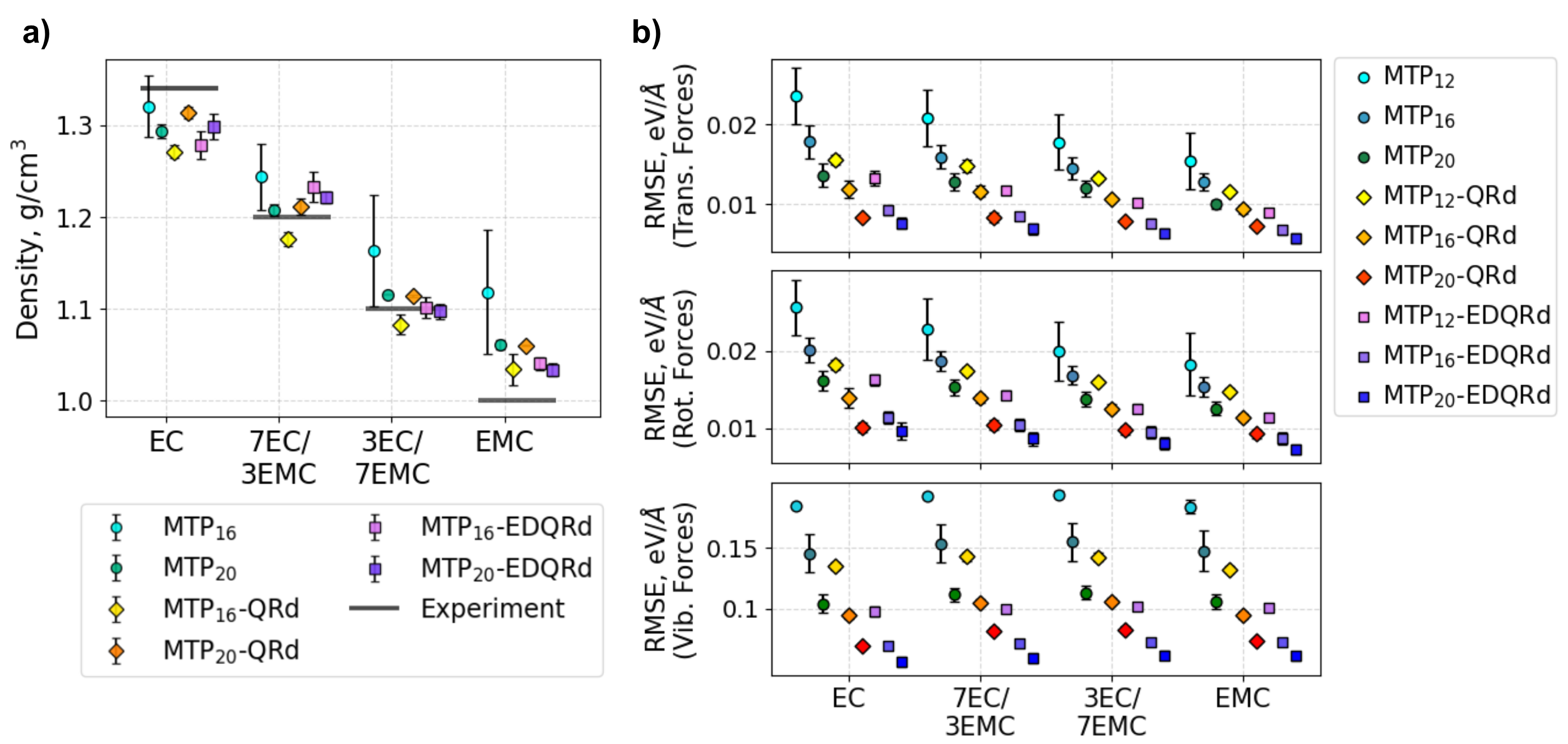}
    \caption{a) Mean densities predicted by an ensemble of three MLIPs with 1-$\sigma$ confidence intervals, compared to literature values~\cite{magduau2023machine} obtained by interpolating experimental data~\cite{johnson1985properties,wang2022current}; Translational, rotational, and vibrational forces, obtained with ensembles of three MTPs, MTP-QRd, and MTP-EDQRd models of levels 12, 16, and 20 (for EDQRd part level 12 was kept for all models) compared with PBE-D3 calculations.}
    \label{fig:mix_ECEMC_densities_ii}
\end{figure*}

To quantify the accuracy difference between all the models, we computed densities for four EC/EMC compositions: pure EC, pure EMC, and 7:3 and 3:7 EC:EMC mixtures (molar ratios). The densities were obtained from the last 100~ps of 300~ps NPT MD simulations performed at 300~K and 1~atm with a 1~fs timestep, initialized from 33–45~\AA\ configurations (see Supplementary Information).
The obtained density values compared with experiment are presented in Fig.~\ref{fig:mix_ECEMC_densities_ii}~(a).
The results demonstrate accuracy improvement due to two independent factors: (i) increasing the MTP level and (ii) incorporating electrostatic interactions via both fixed and environment-dependent charges. For the short-ranged MTP$_{16}$ potential, the mean absolute error lies within 1.5--12\% and increases from pure EC to pure EMC. Moreover, the predicted densities exhibit high uncertainty associated with random initialization of MLIP parameters, leading to a large overlap of the 1-$\sigma$ confidence intervals and indicating an inability to reliably differentiate between different binary solvent compositions.
Upon increasing the MTP level to 20, both issues were largely mitigated, reducing the mean absolute error to 0.6--6\% and the prediction uncertainty to about 1\%.
Notably, incorporating electrostatics via the fixed-charge QRd scheme allows MTP$_{16}$-QRd to achieve accuracy of 1.5--5\%, comparable to that of MTP$_{20}$, while significantly reducing the number of machine-learned parameters. However, further increasing the MTP level within the fixed-charge model (MTP$_{20}$-QRd) does not lead to additional improvement. In contrast, incorporating environment-dependent charges further improves accuracy for MTP$_{20}$-EDQRd compared to all other MLIPs, reducing the average density error to 0.2--3.4\%.

However, density predictions alone do not clearly establish the accuracy ranking among the tested models (e.g., among MTP$_{20}$, MTP-QRd models, and MTP$_{16}$-EDQRd). This stems from the fact that deviations between the predicted and experimental densities reflect both the MLIP fitting error and the intrinsic error of the DFT reference relative to experiment. 
To assess MLIP accuracy relative to the DFT reference, we evaluated the RMSEs of the translational, rotational, and vibrational force components using the methodology and software proposed in Ref.~\cite{magduau2023machine}.
To this end, we constructed a 600-structure validation set by uniformly sampling configurations from 100~ps MD simulations of systems containing 8--16 molecules at four compositions: pure EC, 7EC/3EMC, 3EC/7EMC, and pure EMC. The simulations were performed using MTP$_{16}$, MTP$_{16}$-QRd, and MTP$_{16}$-EDQRd to account for potential differences in configurational-space coverage among the MLIP types and thereby provide an assessment of their accuracy that is not biased toward any single model type.

To assess the effects of both increasing the MTP level and incorporating electrostatics, we employed MTP, MTP-QRd, and MTP-EDQRd models with levels 12, 16, and 20. For each model variant, we trained an ensemble of three MLIPs using the same AL-generated training set obtained for MTP$_{20}$.
Fig.~\ref{fig:mix_ECEMC_densities_ii}~(b) shows the RMSEs for translational, rotational, and vibrational forces on the validation set. Total force and energy errors are reported in the Supplementary Information.

The results demonstrate that increasing the MTP level can partially compensate for the absence of explicit electrostatics, with MTP$_{20}$ achieving accuracy comparable to that of MTP$_{16}$-QRd and MTP$_{12}$-EDQRd. However, incorporating electrostatics additionally improves accuracy: MTP$_{16}$-QRd outperforms MTP$_{20}$, while accounting for environment-dependent charges provides further improvements, allowing MTP$_{16}$-EDQRd to achieve accuracy comparable to that of MTP$_{20}$-QRd.
These results demonstrate that explicit electrostatics enables higher accuracy with fewer machine-learned parameters. For example, MTP$_{16}$-QRd (389 parameters) and MTP$_{12}$-EDQRd (486 parameters) achieve accuracy close to that of MTP$_{20}$ (651 parameters). However, parameter count alone is an important but incomplete measure of computational efficiency; MD simulation speeds and training set sizes are compared in the Supplementary Information.

We also observe that, for all MLIPs, the force RMSE decreases with increasing EMC fraction. Notably, although EMC exhibits larger density errors, it simultaneously shows the smallest deviations in energies and forces from the DFT reference across all EC/EMC compositions. This contrast indicates that density errors reflect not only MLIP fitting accuracy but also the intrinsic error of the DFT reference. Consequently, a reliable ranking of MLIPs emerges only from direct comparison with the \textit{ab initio} data.

\begin{figure*}[!ht]
    \centering
    \includegraphics[width=1\linewidth]{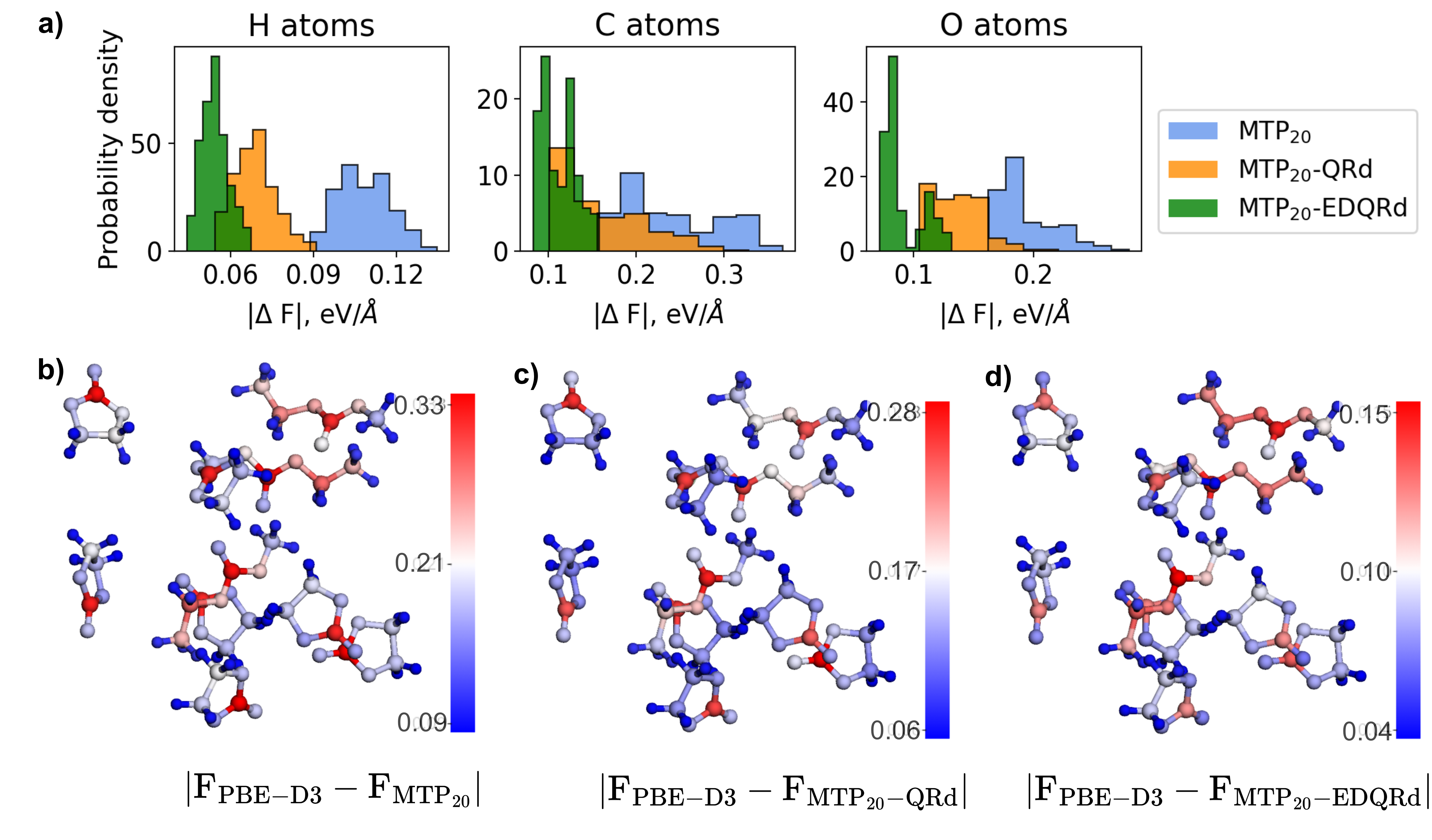}
    \caption{(a) Absolute errors in total forces by atomic type across the full validation set; (b-d) absolute errors in total forces for MTP$_{\rm{20}}$, MTP$_{\rm{20}}$-QRd, and MTP$_{\rm{20}}$-EDQRd, respectively, averaged over the 3EC/7EMC validation configurations. Errors are reported in eV/\AA.}
    \label{fig:ecemc_mix_errors_on_atoms}
\end{figure*}

Before modeling LiPF$_6$ solutions, we assessed MLIP performance by evaluating force errors for each atomic type~(Fig.~\ref{fig:ecemc_mix_errors_on_atoms}). These results demonstrate that incorporating electrostatics through QRd and then EDQRd shifts the error distributions toward lower values, with EDQRd also narrowing the distributions, indicating more consistent accuracy across the evaluated structures (Fig.~\ref{fig:ecemc_mix_errors_on_atoms}~(a)).
Importantly, across all EC/EMC compositions, the MLIPs exhibit relatively small force errors for carbonyl oxygen atoms, which are expected to play a key role in Li$^+$ solvation. Results for the 3EC/7EMC mixture are shown in Fig.~\ref{fig:ecemc_mix_errors_on_atoms}~(b), while results for the other compositions exhibit a similar trend. The inclusion of explicit electrostatics reduces force errors across all atomic types, with a particularly notable improvement for carbonyl carbon: the force error decreases from 0.33~eV/\AA\ for MTP$_{20}$ to 0.15~eV/\AA\ for MTP$_{20}$-EDQRd.

In summary, we demonstrated that AL provides a robust approach to constructing computationally transferable MLIPs of all three types: MTP, MTP-QRd, and MTP-EDQRd. Further analysis showed that incorporating electrostatics improves parameter efficiency, allowing electrostatics-augmented MTPs with fewer parameters to achieve the same or higher force accuracy than short-ranged MTPs, with environment-dependent-charge models achieving the highest accuracy. Interestingly, experimental densities did not clearly rank the models beyond the best and worst performing ones, possibly due to error cancellation between MLIP fitting errors and intrinsic PBE-D3 errors relative to experiment.
Additionally, we verified that all tested MLIPs exhibit moderate force errors on carbonyl oxygen atoms—the primary coordination sites of solvent molecules around Li$^{+}$—suggesting they should reliably capture Li$^{+}$ diffusion. Building on this result, we applied AL to generate training sets for LiPF$_6$ solution MLIPs and then assessed the impact of electrostatic interactions on their accuracy against both \textit{ab initio} and experimental references.

\subsection{Modeling of LiPF$_6$ solution in EC/EMC}

\subsubsection{MLIPs training and training set analysis}

\begin{figure*}[!ht]
    \centering
    \includegraphics[width=1\linewidth]{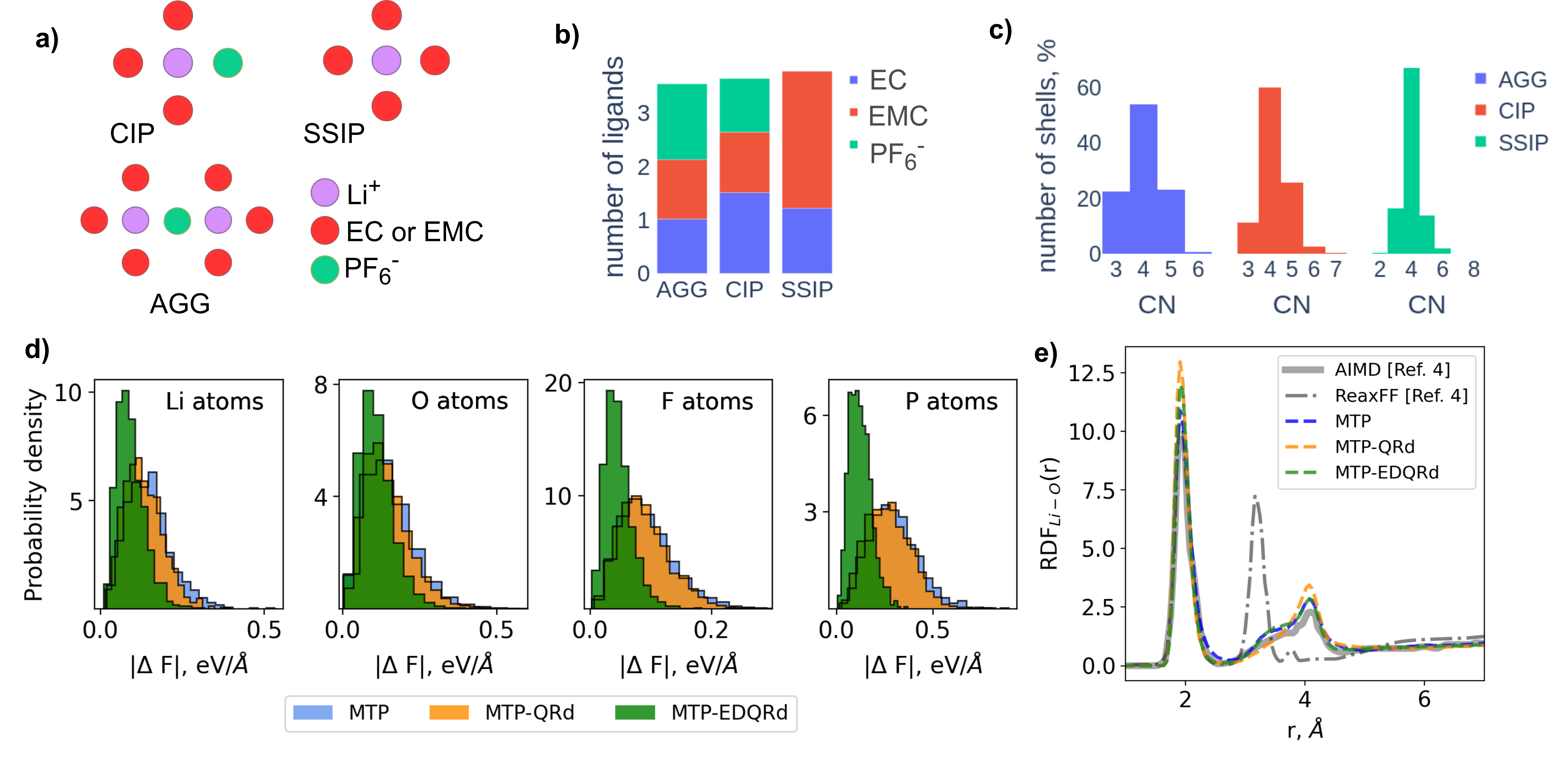}
    \caption{a) Ion pair types; b) number of ligands in Li$^+$ solvation shells present in the training set; c) coordination number (CN) distributions in ion pairs present in training set; d) distributions of force error per atomic type on the validation set; e) radial distribution function (RDF) Li-O for 8 LiPF$_6$ in 80 EC, predicted by MTP compared to the AIMD and ReaxFF ones~\cite{ong2015lithium}.}
    \label{fig:lipf6_train_val}
\end{figure*}

The main challenge in training an MLIP for LiPF$_6$ solutions in EC/EMC arises from the formation of long-lived Li$^+$ solvation shells~\cite{Hou2021}, which hinder AL from comprehensively sampling the relevant configurational space within the accessible AL-MD timescale.
To diversify the configurational space explored in AL, we initialized AL-MD trajectories from representative ion pair (IP) structures: contact ion pairs (CIPs), solvent-separated ion pairs (SSIPs), and aggregates (AGGs) (Fig.~\ref{fig:lipf6_train_val}~(a)).
For each CIP and SSIP type, we constructed four configurations containing 3 EC and 7 EMC molecules, together with one Li$^+$ and one PF$_6^-$ ion. AGG configurations were generated analogously but with twice the number of species.
To select an initial training set, we performed 500~fs AIMD in the NVE ensemble for all IP structures prior to AL.
However, because AGG structures are significantly larger, we adopted a two-stage training strategy to reduce computational cost: we first trained the MLIP on CIP and SSIP data only, and introduced AGGs at a later stage.
For this purpose, we first uniformly sampled ~20 structures from CIP and SSIP AIMD trajectories to train a preliminary MTP. Using this model together with the MaxVol algorithm, we selected additional representative configurations from the same trajectories. The combined uniformly sampled and MaxVol-selected structures formed the CIP/SSIP training set used to train the initial MTP for the subsequent AL. AL-MD simulations were then initiated from CIP and SSIP configurations only.
Once all AL-MD trajectories remained within the reliable extrapolation threshold (see Section \ref{Methods_AL}), we temporarily paused AL and proceeded to a second sampling stage targeting AGG configurations. To this end, we first sampled structures from AGG AIMD trajectories using the MaxVol algorithm, and after that resumed AL with MD initialized exclusively from AGGs.
This two-stage procedure was used to train a level-20 MTP and converged after 86 AL iterations, yielding a training set of 6296 configurations. The large number of iterations and the resulting dataset size highlight the substantially higher complexity of the electrolyte compared to the salt-free solvent, for which the training set contained only 500–1400 configurations.

To assess whether the configurational space was comprehensively sampled, we quantified the diversity of Li$^+$ solvation environments represented in the training set using the SolvationAnalysis module~\cite{cohen2023solvationanalysis} and MDAnalysis package~\cite{gowers2019mdanalysis}. 
We found that the training set primarily consisted of SSIPs (51\%) and CIPs (46\%), with a smaller fraction of AGGs (3\%), which arose from both the focus of most AL iterations on CIPs and SSIPs and the transformation of AGGs into other IP types during AL-MD simulations. Additionally, the observed IP transitions during AL-MD imply that nonequilibrium solvation structures are also present in the training set.
Analysis of ligand coordination around Li$^+$ across IP types (Fig.~\ref{fig:lipf6_train_val}~(b)) indicates the coexistence of both EC-rich and less favorable EMC-rich solvation shells~\cite{ong2015lithium}. The Li$^+$ solvation environments are further characterized by a coordination number (CN) distribution centered around four, consistent with previous studies~\cite{ong2015lithium,franco2019boosting,Hou2021}, while also including both lower- and higher-coordinated species (Fig.~\ref{fig:lipf6_train_val}~(c)). At the finer structural level, we observe both mono- and multidentate coordination of PF$_6^-$ and solvent molecules. Overall, these results support the presence of a diverse range of solvation configurations.
We also note that approximately 3\% of configurations corresponded to unphysical transformations of EC or EMC and were excluded from the analysis, to avoid distorting the evaluation metrics. These configurations were nevertheless retained in the training set to preserve its diversity.

After demonstrating that AL yields a highly diverse training set, we investigated the impact of explicit electrostatic interactions.
In addition to the short-ranged MTP, we applied AL to train MTP-QRd and MTP-EDQRd models with a level-20 MTP component, using the training set previously collected for the local MTP as the starting point. The AL procedure converged within two iterations, requiring only 10--18 additional structures.
Before calculating ionic conductivities, we assessed the accuracy of the MLIPs against the \textit{ab initio} reference. Specifically, we evaluated their ability to predict forces on Li, P, F, and O atoms, which are predominantly involved in solvation interactions, and to reproduce the Li--O radial distribution function (RDF).

\subsubsection{MLIP accuracy with respect to \textit{ab initio} reference}
\label{subsec: mlips_vs_abinitio_lipf6}

To assess the MLIP accuracy in predicting forces, we constructed a validation set of 750 configurations sampled from independent MD trajectories generated using the MTP, MTP-QRd, and MTP-EDQRd models. The trajectories were initialized from 4 SSIP, 4 CIP, and 2 AGG structures and simulated under the same conditions as those used to generate the training set.
The results indicate that incorporating electrostatics through QRd alone yields only a modest improvement in force prediction accuracy, whereas EDQRd provides a more pronounced improvement. The force RMSE decreases only slightly, from 0.085 eV/$\AA$ for MTP to 0.080 eV/$\AA$ for MTP-QRd, but drops to 0.061 eV/$\AA$ for MTP-EDQRd. Consistent with this trend, the force-error norm distributions for Li, O, F, and P atoms (Fig.~\ref{fig:lipf6_train_val}~(d)) show similar means and standard deviations for MTP and MTP-QRd, whereas the MTP-EDQRd distributions are markedly narrower and shifted toward lower errors. The distributions for the remaining atoms are similar across all three models.
These findings suggest that accounting for the dependence of atomic charges on the local environment is particularly important for electrolytes, where charged species and the resulting polarization effects can play a more significant role than in salt-free solvents.

To evaluate how the observed force errors affect structural properties, we calculated the Li--O RDF. We additionally compared the MLIP predictions with the RDFs reported in Ref.~\cite{ong2015lithium}, where both ReaxFF and AIMD simulations were carried out. Notably, this comparison probes the MLIPs in a strongly extrapolative regime, as the RDFs were computed for LiPF$_6$ in pure EC, whereas the models were trained on 3EC/7EMC configurations.
For this test, we calculated the RDF by averaging over a 100~ps NVT MD simulation following prior NPT density equilibration. The simulation cell contained 8~Li$^+$ ions and 80~EC molecules and was initialized from an SSIP configuration, corresponding to the most stable ion pair type in pure EC solvent~\cite{ong2015lithium}.
The RDF predicted by all the tested MLIPs showed excellent agreement with the AIMD results and a clear improvement over ReaxFF~\cite{ong2015lithium}~(Fig.~\ref{fig:lipf6_train_val}~(e)). In particular, MLIPs reproduced the second RDF maximum, associated with coordination by ether oxygen atoms, more accurately than ReaxFF, highlighting its accurate performance even outside the training domain. However, MTP-QRd exhibited limited stability in this extrapolative regime, with simulations lasting only tens of picoseconds on average before failure, despite undergoing the same training procedure as the other MLIPs. Specifically, only several timesteps before the simulation failed, the model predicted unphysically large forces, even though other physical properties, such as density, remained well behaved. This behavior suggests that the model encountered configurations beyond its reliable extrapolation range.

While these results suggest superior extrapolative performance of MTP and MTP-EDQRd, they do not establish whether MTP-QRd is prone to instability near the training domain. Given the comparable force accuracy on validation set of the short-ranged MTP and both electrostatics-augmented models, we proceeded to evaluate ionic conductivity using all three MLIPs.

\subsubsection{Ionic conductivity calculations and trajectories analysis}\label{ICC}

To compute the ionic conductivity $\sigma$, we performed 25-ns equilibrium MD simulations (see Supplementary Information, Section \ref{section:IEC_MD}) for 1M LiPF$_6$ solutions in EC/EMC mixtures with 3:7, 2:3, and 1:1 EC:EMC ratios at 280, 300, and 320~K. These conditions additionally probed the extrapolative performance of the MLIPs, as the training set was generated exclusively from AL-MD simulations at 300~K in the 3EC/7EMC solvent. Under these conditions, the MTP and MTP-EDQRd models produced stable MD trajectories, whereas the fixed-charge MTP-QRd model became unstable even for the 3EC/7EMC system at 300~K, i.e., the closest condition to the training domain. 
Specifically, all MTP-QRd trajectories failed after several hundred picoseconds due to the emergence of unphysically large forces, as observed in the RDF calculations. Therefore, the following results are reported only for the MTP and MTP-EDQRd models.

\begin{figure*}[!ht]
    \centering
    \includegraphics[width=1.0\linewidth]{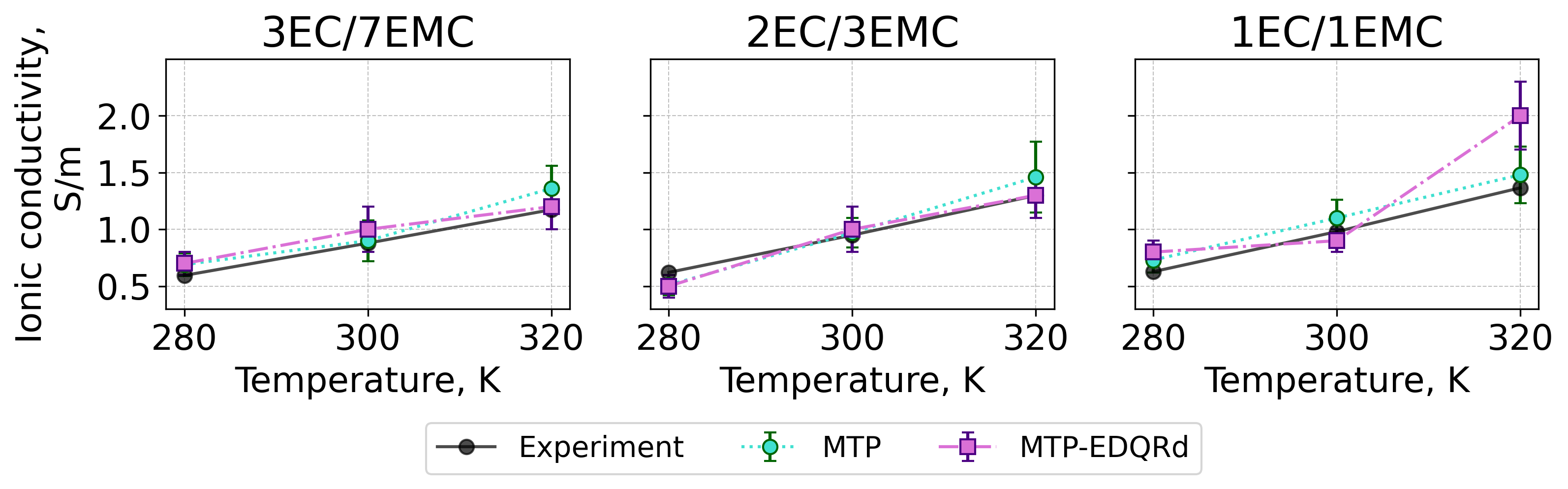}
    \caption{Temperature dependency of ionic conductivity predicted by MTP and MTP-EDQRd, shown together with their uncertainties (see Supplementary Information, Section \ref{section:IEC_MD}) for 1M LiPF$_6$ in solutions at 280, 300 and 320~K. Experimental data were obtained from Ref.~\cite{ding2001change} by sequential quadratic interpolation with respect to temperature and LiPF$_6$ concentration.}
    \label{fig:conductivity}
\end{figure*}

Fig.~\ref{fig:conductivity} compares the computed ionic conductivity with experimental values interpolated over temperature, salt concentration, and solvent composition~\cite{ding2001change}. 
Both MTP and MTP-EDQRd show good agreement with experiment, with mean absolute percentage deviations of 11~\% and 15\%, respectively. 
Furthermore, both models reproduce the experimentally observed increase in ionic conductivity with temperature, consistent with enhanced ion mobility.
However, extrapolation-grade analysis of structures sampled from simulations of 2EC:3EMC and 1EC:1EMC solutions revealed that the MLIP operates in the extrapolation regime (see the Supplementary Information for details). Thus, despite their stability during MD simulations and small deviations from experimental ionic conductivities, the MLIP predictions may not be fully reliable for these compositions and should be interpreted with caution. MLIP retraining may be necessary to improve the fidelity of these simulations, while on-the-fly monitoring of extrapolation grades can be used to detect extrapolative configurations during production simulations.

We additionally assessed finite-size effects by increasing the simulation cell size and found that the resulting changes remain within statistical uncertainty, indicating that a reliable thermodynamic-limit extrapolation would require substantially longer simulations beyond the scope of this work (the details are provided in Supplementary Information).

We further analyzed whether the MTP and MTP-EDQRd model yielded not only similar ionic conductivity predictions, but also electrolyte structures. For this, we analyzed the MD trajectories obtained from ionic conductivity calculations of LiPF$_6$ in 3EC/7EMC at 300~K.
In this analysis, we found that IP composition observed in the MD trajectories produced with MTP and MTP-EDQRd is essentially the same.
While for MTP only SSIPs were found in MD produced and neither CIPs nor AGGs were present, MTP-EDQRd predicted the presence of only 2.5\% CIPs alongside SSIPs, indicating overall agreement in predicting near-complete salt dissociation.
Experimental and computational studies indicate that LiPF$_6$ in pure EC contains only a small fraction of CIPs~\cite{haghkhah2020effect, smirnov2019structure}. Increasing the fraction of low-dielectric solvents (e.g., EMC) promotes ion pairing, with CIPs observed experimentally in 1M LiPF$_6$ in 1EC/9EMC~\cite{feng2019communication} and reported in simulations of 3EC/7EMC using classical force fields~\cite{ringsby2021transport}. However, nonpolarizable force fields are known to overestimate ion pairing~\cite{kasemchainan2025enhanced}, and systematic experimental data on CIP concentrations as a function of EC/EMC ratio remain limited.

\begin{figure}[!ht]
    \centering
    \includegraphics[width=1\linewidth]{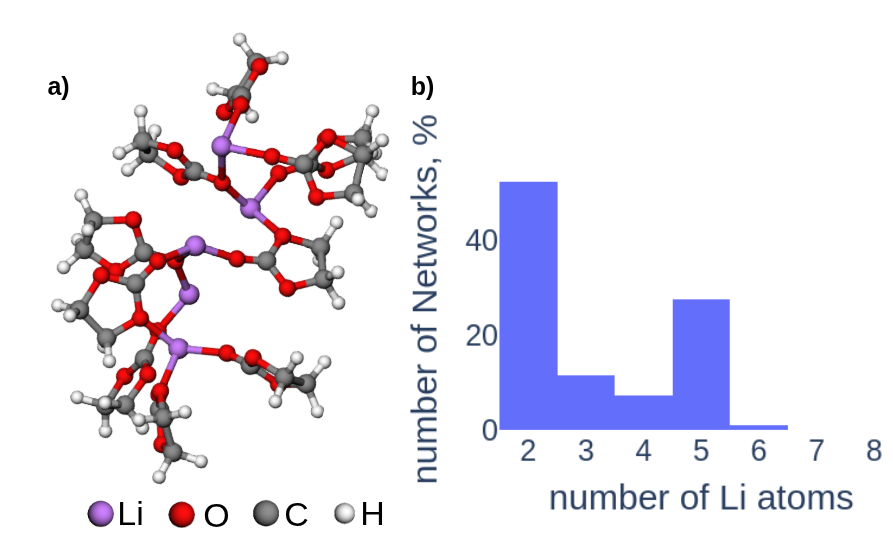}
    \caption{a) Interconnected SSIPs (Network) found in MD produced by MTP, with only EC molecules being shown; b) distribution of number of Li$^+$ ions involved into Networks, encountered in MD produced by MTP.}
    \label{fig:long_trj_analysis}
\end{figure}

Additionally, in MD simulations performed with MTP, we found that approximately 85\% of SSIPs form interconnected networks linked via shared EC molecules, with network sizes ranging from 2 to 8 Li$^+$ ions (Fig.~\ref{fig:long_trj_analysis}~(b)). Within these networks, EC molecules act as bridges between Li$^+$ ions via coordination through both carbonyl and ether oxygen atoms (Fig.~\ref{fig:long_trj_analysis}~(a)).
Li$^+$ ions participating in networks exhibit a higher EC content in their solvation shells compared to isolated SSIPs: while maintaining an average coordination number close to four, networked Li$^+$ ions show a shift toward higher coordination, with an increased fraction of CN = 5 and a reduced fraction of CN = 3 (see Supplementary Information).
Similar configurations were observed in MTP-EDQRd MD simulations, comprising 10\% of all Li$^+$ solvation shells, although only two interconnected solvation shells were present in these structures. Interestingly, these structural differences did not have a pronounced effect on ionic conductivities. Given that these structural differences pose a challenge for validation against experiment or ab initio simulations, we treat them as the observation of MLIP behavior difference, not as physical insight on the electrolyte structure.

In summary, both MTP and MTP-EDQRd accurately reproduce ionic conductivity and its temperature dependence compared to experiment. Interestingly, both models show similar accuracy relative to experiment despite a pronounced difference in accuracy relative to the \textit{ab initio} reference (see Section~\ref{subsec: mlips_vs_abinitio_lipf6} for details). Additionally, MTP-EDQRd outperformed the fixed-charge MTP-QRd model in terms of MD stability. Further assessment of whether incorporating Gaussian charges, as in the EDQRd model, could improve the stability of MTP-QRd showed no improvement, highlighting the importance of incorporating environment-dependent charges (details are provided in the Supplementary Information).

\section{Conclusions}

In this study, we demonstrated that active learning (AL) based on the D-optimality criterion and the MaxVol algorithm~\cite{podryabinkin2017active} enables largely automated generation of diverse training sets for MLIPs, requiring manual setup primarily of the initial structures and target simulation conditions. Importantly, we extended this AL framework to long-range electrostatics-augmented MTPs, addressing the previously unavailability of AL for these models.
We applied this approach to pure EC and EMC, EC/EMC mixtures, and LiPF$_6$ solutions in EC/EMC, obtaining models that maintain stable densities in molecular dynamics (MD) simulations and transfer across solvent compositions.

For the EC/EMC binary solvent, electrostatics-augmented MLIPs achieved accuracy comparable to or better than that of short-ranged MTPs with fewer parameters. For LiPF$_6$ electrolyte solutions, however, a notable improvement over the short-ranged MTP was observed only for MTP-EDQRd. Additionally, MTP-EDQRd outperformed MTP-QRd in terms of MD stability for electrolyte solutions, enabling stable MD simulations for various EC/EMC ratios at the scale of thousands of atoms and dozens of nanoseconds, matching MTP in stability. The resulting MTP and MTP-EDQRd models reproduced experimental densities of EC, EMC, and their mixtures with mean absolute deviations of 0.2--6\%, and ionic conductivities with deviations of 11--15\%, while capturing the experimentally observed increase in conductivity with temperature.

In summary, our results demonstrate that AL enables highly automated generation of diverse training sets for MLIPs. At the same time, explicitly incorporating electrostatic interactions offers a promising route to improving parameter efficiency of MTPs while maintaining or enhancing accuracy. Although the fixed-charge QRd model represents an initial step toward this goal, the environment-dependent-charge EDQRd model provides more substantial improvements in accuracy and MD stability. Future work will focus on assessing the accuracy of EDQRd for systems with different total-charge states and on developing models with charges that depend on the global atomic environment, aiming to further improve accuracy and extend the applicability of electrostatics-augmented MLIPs.

\section{Data availability}

The training sets, trained potentials, and VASP and LAMMPS simulation settings will be made publicly available at GitHub repository https://github.com/OlgaChalykh/ mlips\_for\_molecular\_liquids\_and\_electrolytes.

\section{Code availability}

The MLIP-2 code used for AL and training of MTP potentials is available at GitLab repository https://gitlab.com/ashapeev/mlip-2.

The MLIP-4 code used for AL, training of MTP-QRd and MTP-EDQRd potentials is available at GitLab repository https://gitlab.com/ashapeev/mlip-4.

\section{Funding}
This study was supported by Russian Science Foundation (grant number 23-13-00332-P, https://rscf.ru/project/23-13-00332/).

\section{Authors contributions}
O.C. and N.R. conceptualized the project. N.R. performed the initial calculations. O.C. and D.K. trained the potentials, and O.C. carried out their validation and the analysis of results. M.P. performed the ionic conductivity calculations. D.K. extended active learning to electrostatics-augmented potentials. O.C. drafted the initial manuscript, and M.P., D.K., N.R., and A.S. contributed to writing and refining the final version. N.R. and A.S. supervised the study.

\section{Competing interests}
The authors declare no competing interests.

\newpage
%%%%%%%%%% Prefix a "S" to all equations, figures, tables and reset the counter %%%%%%%%%%
\setcounter{equation}{0}
\setcounter{figure}{0}
\setcounter{table}{0}
\setcounter{section}{0}
\makeatletter
\renewcommand{\theequation}{S\arabic{equation}}
\renewcommand{\thefigure}{S\arabic{figure}}
\renewcommand{\thesection}{S\arabic{section}}
\renewcommand{\thesubsection}{\thesection.\arabic{subsection}}

\onecolumn

\section*{Supplementary Information}

This Supplementary Information provides additional methodological details and supporting analyses for the main text. It outlines the mathematical formulations for the Moment Tensor Potential (MTP), fitting procedures, the active learning protocol, and the Green-Kubo formalism for ionic conductivity calculations. Furthermore, we describe the training pipeline for compositionally transferable potentials, present density evolution over time, and tabulate the MLIP-predicted density values. Finally, we report results on ionic conductivity convergence, finite-size effects, the fidelity of calculations performed outside the training domain, and additional structural analyses of the electrolyte structure.

\section{Methodology}
\subsection{Moment Tensor Potential}
\label{methods_mtp}

MTP energy is represented as a sum of $N$ contributions $V^{\rm MTP}(\mathfrak{\bm n}_i)$, corresponding to energy of atom $i$ in its neighborhood ${\bf \mathfrak{n}}_i$:
\begin{align} \label{EnergyMTP}
E^{\rm MTP} = \sum \limits_{i=1}^{N} V^{\rm MTP}(\mathfrak{\bm n}_i),
\end{align} 
where each neighborhood is limited by a cutoff radius $R_{\rm cut}$ and comprises information on atomic types of central and neighboring atoms $z_i$, $z_j$ and their relative positions r$_{ij}$:
$$\mathfrak{ n}_i = ( \{r_{ij}<R_{\rm cut},z_i,z_j \}_{j=\overline{1,\;N_{\rm nbh}}} ),$$
where $N_ {\rm nbh}$ is the neighborhood size.

Each contribution $V^{\rm MTP}(\mathfrak{\bm n}_i)$ is expanded over a set of MTP basis functions $B_{\alpha}$: 
\begin{align} \label{SiteEnergyMTP}
V^{\rm MTP}({\bf \mathfrak{n}}_i) = \sum \limits_{\alpha} \xi_{\alpha} B_{\alpha}({\mathfrak{\bm n}}_i),
\end{align} 
where $\xi_{\alpha}$ are the linear parameters to be optimized during potential fitting. 

The set of MTP basis functions depends on a parameter called level of MTP, or the maximum level, ${\rm lev_{\rm max}}$. Only such functions are included in the basis set of MTP that ${\rm lev} B_{\alpha} \leq {\rm lev_{\rm max}}$. The level of a basis function is defined as follows:
\begin{equation} \label{LevelMultMTD}
\displaystyle
{\rm lev} B_{\alpha} = \rm {lev} \prod_{p=1}^{P} M_{\mu_p,\nu_p},
\end{equation}
where $M_{\mu,\nu}$ refers to moment tensor descriptor, expressed as a product of the angular $r_{ij}^{\otimes \nu}$ (the symbol ``$\otimes$'' denotes the outer product of vectors) and radial part $f_{\mu}(|r_{ij}|,z_i,z_j)$:  

\begin{equation}\label{MomentTesnsorDescriptors}
M_{\mu,\nu}({\mathfrak{\bm n}}_i)=\sum_{j=1}^{N_{\rm nbh}} f_{\mu}(|r_{ij}|,z_i,z_j) r_{ij}^{\otimes \nu}.
\end{equation}
The radial part $f_{\mu}(|r_{ij}|,z_i,z_j)$ follows a form
\begin{align} \label{RadialFunction}
\displaystyle
f_{\mu}(|r_{ij}|,z_i,z_j) = \sum_{\beta} c^{(\beta)}_{\mu, z_i, z_j} T^{(\beta)} (|r_{ij}|) (R_{\rm cut} - |r_{ij}|)^2.
\end{align}
Here $\mu$ is the number of the radial function $f_{\mu}$, $c=\{c^{(\beta)}_{\mu, z_i, z_j}\}$ are the radial parameters to be found in potential fitting, $T^{(\beta)} (|r_{ij}|)$ are polynomial basis functions, and the term $(R_{\rm cut} - |r_{ij}|)^2$ is introduced to ensure smoothness with respect to the atoms leaving and entering the sphere with the cutoff radius $R_{\rm cut}$.

Thus, the level of $M_{\mu,\nu}$ is defined as:
\begin{equation} \label{eq:LevelMTD}
\displaystyle
{\rm lev} M_{\mu,\nu} = 2 + 4 \mu + \nu.
\end{equation}

\subsection{Ionic conductivity calculations}
\label{section:IEC_MD}

For ionic conductivity calculations, we employ the Green-Kubo formalism~\cite{GK_Green1954, GK_Kubo1957} to account for all the correlations in ions' motion~\cite{Kubisiak2020, Verma2024}, as we expect to observe various stable ion pairs and complexes in solution. The Green-Kubo method relates the charge current in the simulation cell to ionic conductivity via the time integral of the autocorrelation function, given by:
\begin{equation}
\label{eq:sigma}
    \sigma = \frac{1}{3Vk_bT}\int\limits_0^{\infty} \langle {\bf J}(t), {\bf J}(0) \rangle dt, 
\end{equation}
where $\sigma$ is ionic conductivity, $V$ -- volume of the system, $k_b$ -- Boltzmann constant, $T$ -- temperature in the system. ${\bf J}(t)$ -- is a charge current, that is calculated by the following expression:
\begin{equation}
\label{eq:J}
    {\bf J}(t) = \sum_{i=1}^{N_{ch}} q_i{\bf v}_i(t),
\end{equation}
where $q_i$ and ${\bf v}_i(t)$ are the charge and velocity of the $i$-th charged particle, $N_{ch}$ is the number of charged particles accounted. Calculating the ionic conductivity, we assume that charge in the LiPF$_6$ EC/EMC electrolyte is transferred only by Li$^+$ and PF$_6^-$ ions. This means that we do not include solvent molecules (EC and EMC) in the calculations of the charge current in Eq.~\ref{eq:J}, as they may introduce additional noise in the autocorrelation function in Eq.~\ref{eq:sigma}, increasing the uncertainty of the ionic conductivity calculations. The charge of Li$^+$ is set to +1. Calculating the contribution of [PF$_6$]$^-$ to the charge current we only use the velocity of the P atom, as its position corresponds to center of mass of complex ion, and set the charge of [PF$_6$]$^-$ ion to -1. 

A well-known issue in Green-Kubo calculations of transport properties is the convergence of the integral in Eq.~\ref{eq:sigma}~\cite{maginn2019best}. Since infinitely long simulations are impossible, the upper limit of the integral is replaced by a finite correlation time $\tau_{\text{c}}$ (i.e., the integral is truncated at $\tau_{\text{c}}$). To obtain accurate conductivity values, the correlation time should be sufficiently long for the autocorrelation function $\langle{\bf J}(\tau_{\text{c}}), {\bf J}(0) \rangle$ to decay to zero.
In our calculations, we first performed MD simulations with a correlation time of 1~ns to verify convergence. We then examined the dependence of ionic conductivity $\sigma$ on the correlation time $\tau_{\text{c}}$ and selected a correlation time at which the further increase of $\tau_{\text{c}}$ did not change the conductivity value. Accordingly, we set $\tau_{\text{c}}$ to 500~ps for MTP and MTP-EDQRd simulations (all dependencies of ionic conductivity $\sigma$ on time are presented in Supplementary Information).
However, even in the converged part of the integral, $\sigma(t)$ exhibited oscillations. To mitigate their effect on the calculated ionic conductivity, we averaged $\sigma(t)$ over both independent trajectories and time within the converged region. % of $\sigma(t)$ dependence
Thus, for each system studied, several autocorrelation function integrals were obtained in independent trajectories, and then the average autocorrelation function integral was calculated.
For every point in $\sigma(t)$, we obtained the uncertainty $\delta\sigma(t)$ as the standard error of the mean over the independent trajectories. 
To eliminate the oscillations in $\sigma$(t), we averaged the $\sigma$ value in the converged part of the integral over the last 300~ps. The reported uncertainties in our calculations correspond to the mean of $\delta\sigma(t)$ values, representing a conservative estimate of the uncertainty, since it assumes that all errors $\delta\sigma(t)$ are correlated. 

In this study, we calculated ionic conductivity using MTP and MTP-EDQRd for a 1M LiPF$_6$ solution in a 3EC/7EMC (molar ratio) binary solvent at 1~atm and 280, 300, and 320~K. The simulations employed cells containing 8 Li$^+$ and PF$_6^-$ ions, along with the corresponding number of solvent molecules: 24/56 EC/EMC molecules for 3:7 ratio, 32/48 EC/EMC molecules for 2:3 ratio, and 40/40 EC/EMC molecules for 1:1 ratio. The simulation cell dimensions were approximately 23~\AA.

For the ionic conductivity calculations, we followed the workflow described in Ref.~\cite{Hou2021} to equilibrate the simulation cell and obtain the equilibrium distribution of ion pair types. The equilibration procedure consisted of a series of NPT simulations: 1~ns at 300~K, followed by 1~ns of heating to 380~K and 1~ns at 380~K, then 1~ns of cooling back to 300~K and a final 1~ns equilibration at 300~K. The resulting structure was used as the starting point for ionic conductivity calculations.
To generate independent trajectories from the equilibrated structure, atom velocities were reassigned according to the Maxwell distribution using different random seeds. Prior to the production runs, each trajectory was equilibrated in the NPT ensemble for 500~ps. In total, we ran 10 independent trajectories of 2.5~ns each, corresponding to 25~ns overall simulation time per temperature for both MTP and MTP-EDQRd models.

% \textcolor{red}{To diffusion coefficient of species are dependent on the size of simulation cell in MD. The ionic conductivity should exhibit the same behavior. To check, whether we can eliminate this we conduct additional modeling for bigger cells containing about 2000 atoms in total with the box size of 28 \AA.}

\section{Modeling of EC/EMC mixtures}

\subsection{Training of compositionally transferable potential}

\begin{figure}[!ht]
    \centering
    \includegraphics[width=1\linewidth]{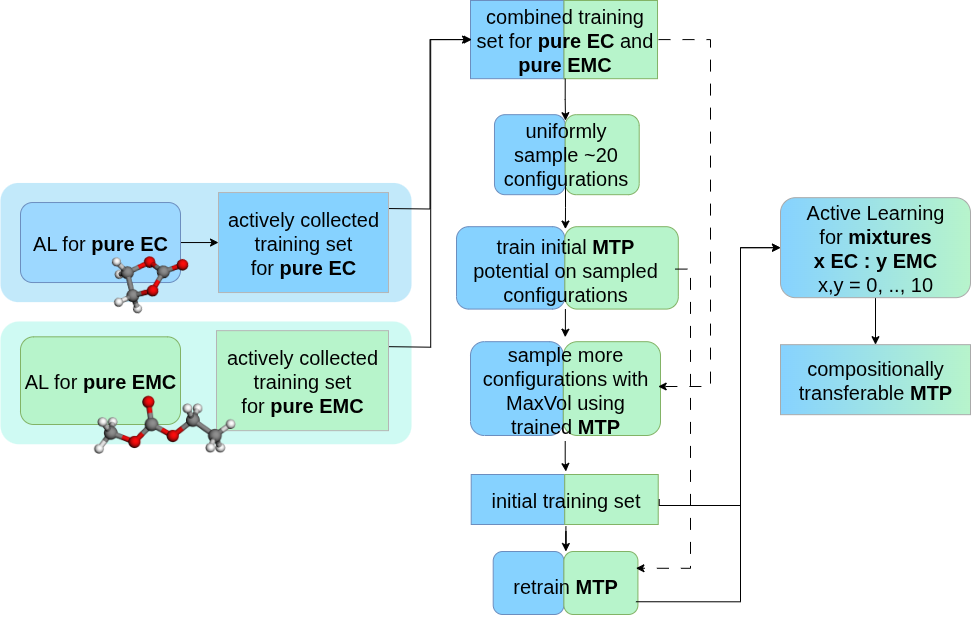}
    \caption{The pipeline proposed for training the compositionally transferable MTP.}
    \label{fig:ct_training_pipeline}
\end{figure}

The pipeline purposed for training the compositionally transferable MTP is illustrated in Figure \ref{fig:ct_training_pipeline}. This pipeline takes as input the training sets collected for pure EC and EMC and uses them to generate a pretraining set. First, the training sets of pure EC and EMC are combined, and approximately 20 configurations are uniformly sampled to train an initial MTP. Second, the resulting MTP is then used to select additional configurations from the combined EC and EMC training set  using MaxVol algorithm \cite{goreinov2010-maxvol}. After that, the uniformly sampled and MaxVol-selected configurations together form the initial training set, which is used to retrain the initial MTP. Finally, the initial training set and retrained MTP are passed to the AL\cite{novikov2020mlip}, where new configurations are sampled from AL-MD simulations initialized at configurations spanning the full compositional space of EC/EMC mixtures.

When the described pipeline applied to level-20 MTP, it yields a training set with the following composition: 35.7\% EC, 56.9\% EMC, and 7.3\% EC/EMC mixtures.
% , where xx\% of the EC and xx\% of the EMC configurations are inherited from the initial training set.
% Pretraining set: 41.6\% EC, 58.4\% EMC; training set: ; 

\clearpage
After the training sets for the MTP potentials were generated, we used them to initialize active learning (AL) for the MTP-QRd and MTP-EDQRd models. The inventory table summarizing the resulting models' names, corresponding training set sizes, approximate training costs, and other details is provided in Table \ref{tab:SI_mlip_efficiency}. We also note that assessment of MD simulation speed was performed in MLIP-4~\cite{mlip4} software for all MLIPs. Additionally, the fitting weights are provided in Section 2.3 of the manuscript.

\begin{table*}[ht]
    \centering
    \caption{The inventory table summarizing the architecture and computational efficiency of MLIPs used for EC/EMC mixtures simulation.}
    \renewcommand{\arraystretch}{1.3}
    \begin{tabular}{|p{3.2cm}|p{0.8cm}|c|c|c|c|c|c|}
        \hline
        \multicolumn{2}{|p{2cm}|}{\textbf{MLIP type}} & \multicolumn{2}{c|}{\textbf{MTP}}
        & \multicolumn{2}{c|}{\textbf{MTP-QRd}}
        & \multicolumn{2}{c|}{\textbf{MTP-EDQRd}} \\
        \cline{1-8}
        \multicolumn{2}{|p{2cm}|}{\textbf{MTP name}}
        & MTP$_{16}$ & MTP$_{20}$
        & MTP$_{16}$-QRd & MTP$_{20}$-QRd
        & MTP$_{16}$-EDQRd & MTP$_{20}$-EDQRd\\
        \cline{1-8}
        \multicolumn{2}{|p{2cm}|}{\textbf{MTP level}}
        & 16 & 20
        & 16 & 20
        & 16 & 20 \\
        \hline
        \multicolumn{2}{|p{4cm}|}{\textbf{Number of parameters}}
        & 383 & 651 & 389 & 657 & 621 & 889 \\ \hline
        \multicolumn{2}{|p{4cm}|}{\textbf{Average training set size}}
        & 517 & 530 & 500 & 1031 & 719 & 1527 \\ \hline
        \multicolumn{2}{|p{4cm}|}{\textbf{Approximate MLIP training cost (CPU hours)}}
        & 21 & 53 & 40 & 76 & 57 & 210 \\ \hline
        \multirow{2}{*}{\parbox{3.2cm}{\textbf{Force-evaluation throughput}\\
        (MD simulation speed, ns/day on 16 CPUs)}}
        & 160 atoms & 182.1 & 79.9 & 81.5 & 52.3 & 44.6 & 33.5 \\ \cline{2-8}
        & 4320 atoms & 9.5 & 4.1 & 1.6 & 1.5 & 1.6 & 1.4 \\
        \hline
    \end{tabular}
    \label{tab:SI_mlip_efficiency}
\end{table*}

\clearpage
\subsection{EC/EMC mixtures density calculations}

\begin{figure}[!ht]
    % \centering
    \includegraphics[width=1\linewidth]{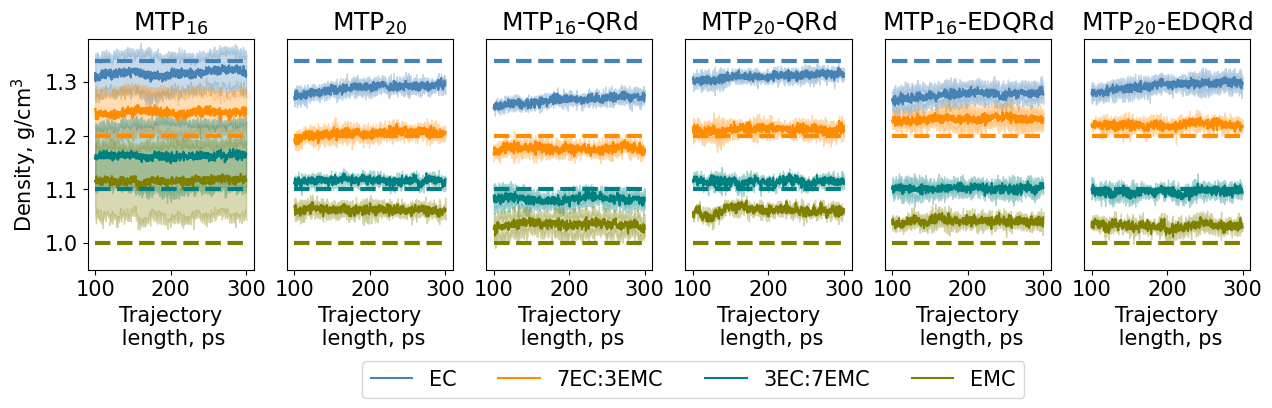}
    \caption{
    Evolution of density over time obtained with MTP$_{16}$, MTP$_{20}$, MTP$_{16}$-QRd, MTP$_{20}$-QRd, MTP$_{16}$-EDQRd, and MTP$_{20}$-EDQRd}.
    The MTP models were actively trained, while the MTP-QRd and MTP-EDQRd models were trained on the dataset actively collected for MTP$_{20}$. The experimental value is depicted by a dashed line, the predicted value by a solid line, and the semi-transparent interval represents the standard deviation over an ensemble of three MLIPs.
    \label{fig:ct_rho_over_t}
\end{figure}

When all the models listed in Table~\ref{tab:SI_mlip_efficiency}, we proceeded to calculate the densities. To complement the mean density analysis presented in the main text, we additionally report the full time evolution of the density during molecular dynamics (MD) simulations (Figure~\ref{fig:ct_rho_over_t}).

Additionally, we provide the density values reported in the main text, averaged over the last 100~ps of these simulations, in tabular format (Table~\ref{tab:SI_density_values}). The table includes both the mean and standard deviation to quantify the uncertainty arising from random MLIP parameter initialization.

\begin{table*}[ht]
    \centering
    \renewcommand{\arraystretch}{1.3}
    \begin{tabular}{|c|c|c|c|c|}
        \hline
         & EC & 7EC:3EMC & 3EC:7EMC & EMC  \\ \hline
        MTP$_{16}$ & 1.32 $\pm$ 0.03 & 1.24 $\pm$ 0.04 & 1.16 $\pm$ 0.06 & 1.12 $\pm$ 0.07 \\ \hline  
        MTP$_{20}$ & 1.29 $\pm$ 0.01 & 1.21 $\pm$ 0.01 & 1.12 $\pm$ 0.0 & 1.06 $\pm$ 0.0 \\ \hline  
        MTP$_{16}$-QRd & 1.27 $\pm$ 0.01 & 1.18 $\pm$ 0.01 & 1.08 $\pm$ 0.0 & 1.03 $\pm$ 0.02 \\ \hline
        MTP$_{20}$-QRd & 1.31 $\pm$ 0.01 & 1.21 $\pm$ 0.01 & 1.11 $\pm$ 0.0 & 1.06 $\pm$ 0.0 \\ \hline
        MTP$_{16}$-EDQRd & 1.28 $\pm$ 0.02 & 1.23 $\pm$ 0.02 & 1.10 $\pm$ 0.01 & 1.04 $\pm$ 0.01 \\ \hline
        MTP$_{20}$-EDQRd & 1.30 $\pm$ 0.01 & 1.22 $\pm$ 0.02 & 1.10 $\pm$ 0.01 & 1.03 $\pm$ 0.01 \\ \hline
    \end{tabular}
    \label{tab:SI_density_values}
\end{table*}

\clearpage
\subsection{Energy and force prediction errors}

\begin{figure}[!ht]
    \centering
    \includegraphics[width=0.8\linewidth]{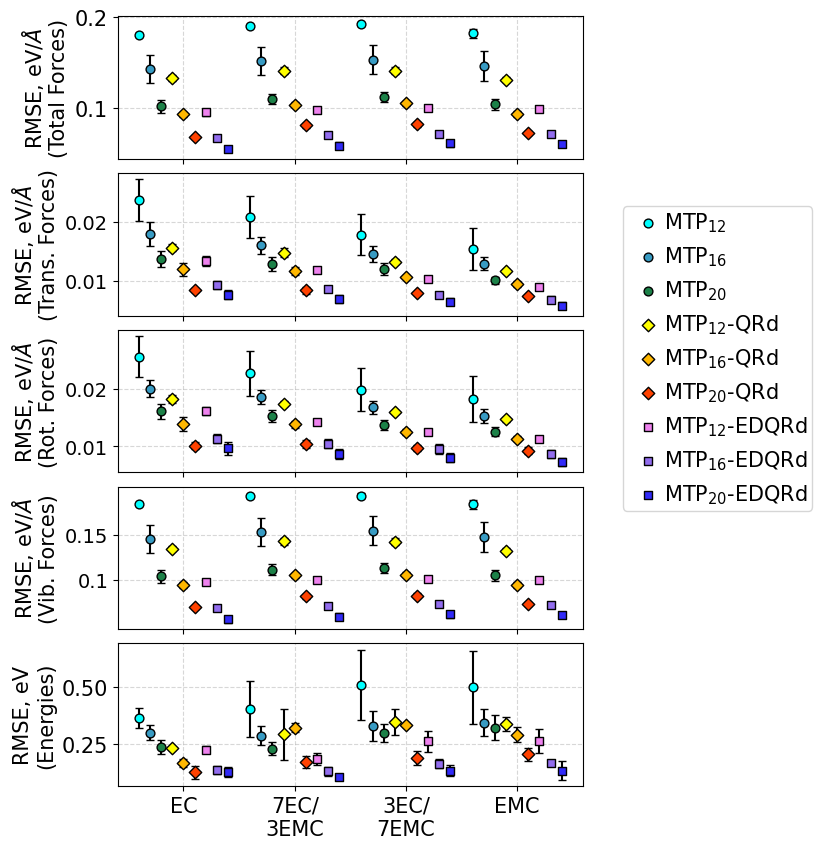}
    \caption{
    RMSEs of energies and total, translational, rotational, and vibrational forces for ensembles of three MTP, MTP-QRd, and MTP-EDQRd potentials. Both mean and standard deviations are presented to quantify the uncertainty arising from random initialization of MLIP parameters.}
    \label{fig:SI_mix_ecemc_ii}
\end{figure}

Fig.~\ref{fig:SI_mix_ecemc_ii} shows the energy errors and the total, translational, rotational, and vibrational force errors. The results demonstrate that both energy and force errors decrease upon (i) increasing the MTP level and (ii) introducing explicit electrostatics, with the MTP-EDQRd models achieving the highest accuracy.

\clearpage

\begin{figure}[!ht]
    \centering
    \includegraphics[width=0.9\linewidth]{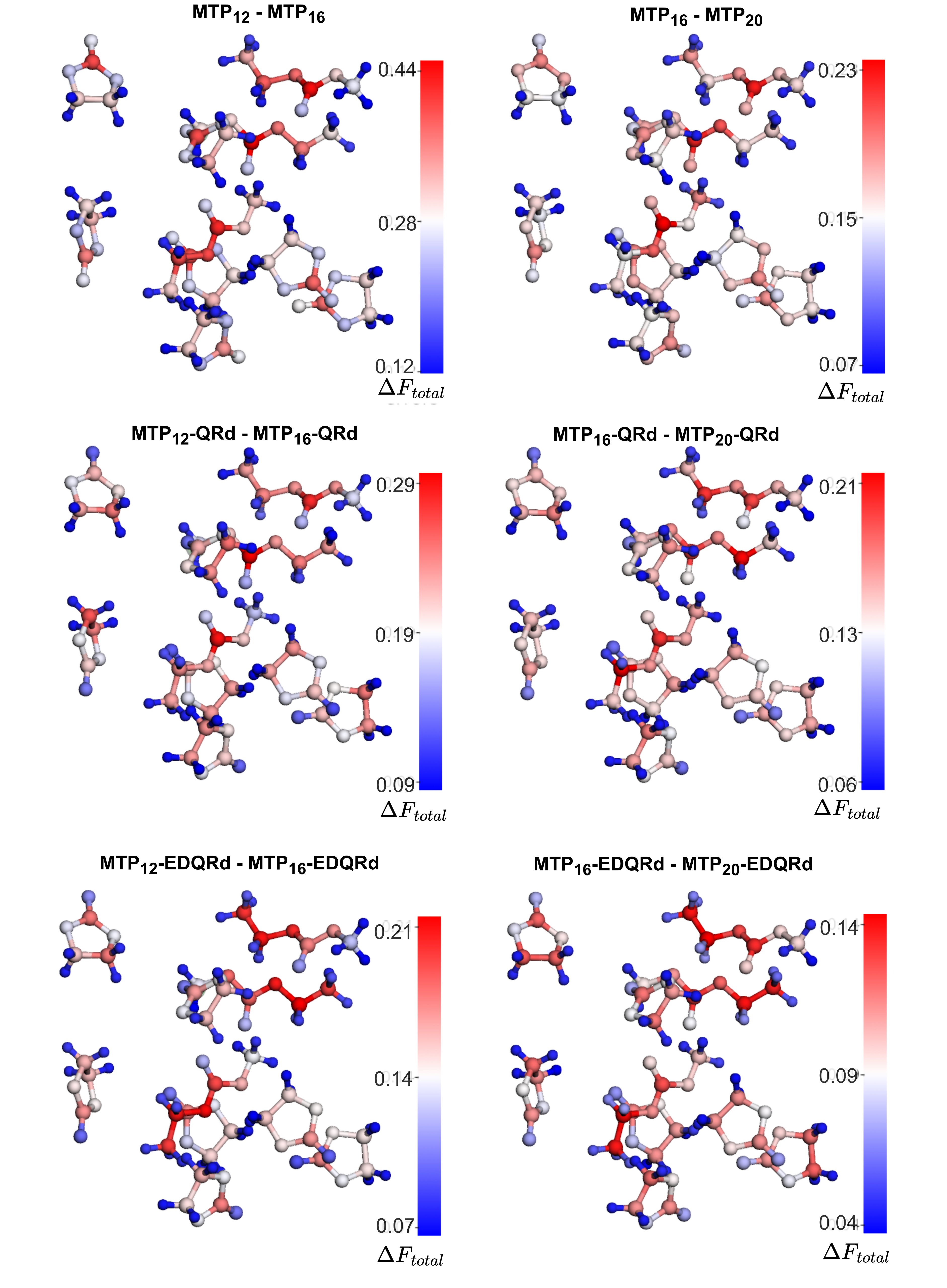}
    % \caption{}
    \caption{
    Difference in total forces between MTP, MTP-QRd, and MTP-EDQRd models with MTP parts of levels 12, 16, and 20, all trained on the dataset actively collected for MTP$_{20}$}.
    \label{fig:ct_Ferrors_diff}
\end{figure}

Figure~\ref{fig:ct_Ferrors_diff} presents the difference in forces between MTP, MTP-QRd, and MTP-EDQRd models with MTP parts of levels 12, 16, and 20, all trained on the dataset actively collected for MTP$_{20}$.

\begin{figure}[!ht]
    \centering
    \includegraphics[width=1\linewidth]{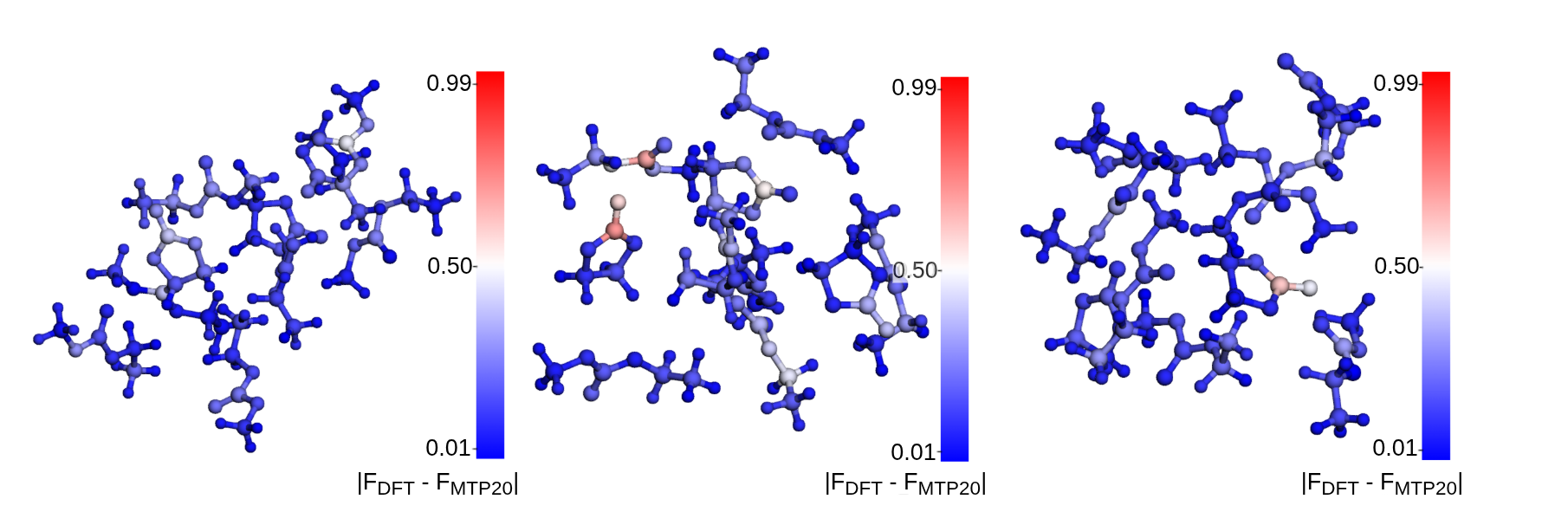}
    % \caption{}
    \caption{
    Force RMSE for individual MD snapshots predicted by MTP$_{20}$.
    }
    \label{fig:ct_Ferrors_snapshots}
\end{figure}

Figure \ref{fig:ct_Ferrors_snapshots} shows force RMSE predicted by MTP$_{20}$ for individual MD snapshots, without averaging over configurations. The figure illustrates that, although RMSE is on average concentrated on carbonyl carbons, not all of these atoms are affected by large errors in particular snapshots.

\subsection{Stress prediction errors}

To achieve accurate densities in NPT simulations, the MLIP must also accurately reproduce stresses. Since current training sets used for MLIP training omit this information, stresses are inferred from energies and forces. Therefore, we verified the accuracy of the predicted stresses. To this end, we evaluated the MTP${20}$, MTP$_{20}$-QRd, and MTP$_{20}$-EDQRd models on a validation set for EC/EMC mixtures. The resulting errors for all stress components, reported in GPa in the table below (Figure~\ref{tab:SI_pred_stresses}), indicate that the stresses are inferred with sufficient accuracy, comparable to those reported in other studies employing MLIPs~\cite{wang2022lattice,chen2022universal}.

\begin{table*}[ht]
    \centering
    \renewcommand{\arraystretch}{1.3}
    \begin{tabular}{|c|c|c|c|}
        \hline
        & MTP$_{20}$ & MTP$_{20}$-QRd & MTP$_{20}$-EDQRd\\ \hline
        RMSE & 0.486 & 0.490 & 0.489 \\ \hline
        MAE & 0.323 & 0.323 & 0.326 \\ \hline
    \end{tabular}
    \label{tab:SI_pred_stresses}
\end{table*}

\clearpage

\section{Modeling of LiPF$_6$ solution in EC/EMC}

\subsection{Locality test}
To validate the 5~\AA{} cutoff for LiPF$_6$ in EC/EMC, we followed the procedure described in Ref.~\cite{magduau2023machine}. After converging the density using an actively trained MTP in a 10~\AA{} cell, we froze all atoms within 5~\AA{} of the Li atom and performed a 100~ps MD simulation. Ten structures were sampled from this simulation and recalculated using \textit{ab initio} settings identical to both those used for our training set and those employed in Ref.~\cite{magduau2023machine}. The standard deviation of the force magnitude on the Li atom was 0.05~eV/\AA{}, which is below the 0.1~eV/\AA{} threshold established for EC/EMC solvents in Ref.~\cite{magduau2023machine} and comparable to the average MTP force fitting error (0.04~eV/\AA{}) on the actively collected dataset for the LiPF$_6$/EC/EMC solution. These results confirm that the 5~\AA{} cutoff enables the MTP to accurately capture Li-solvation interactions, which are essential for calculating ionic conductivity.

\clearpage

\subsection{Ionic conductivity calculations}

\subsubsection{Ionic conductivity convergence analysis}

\begin{figure}[!ht]
    \centering
    \includegraphics[width=0.8\linewidth]{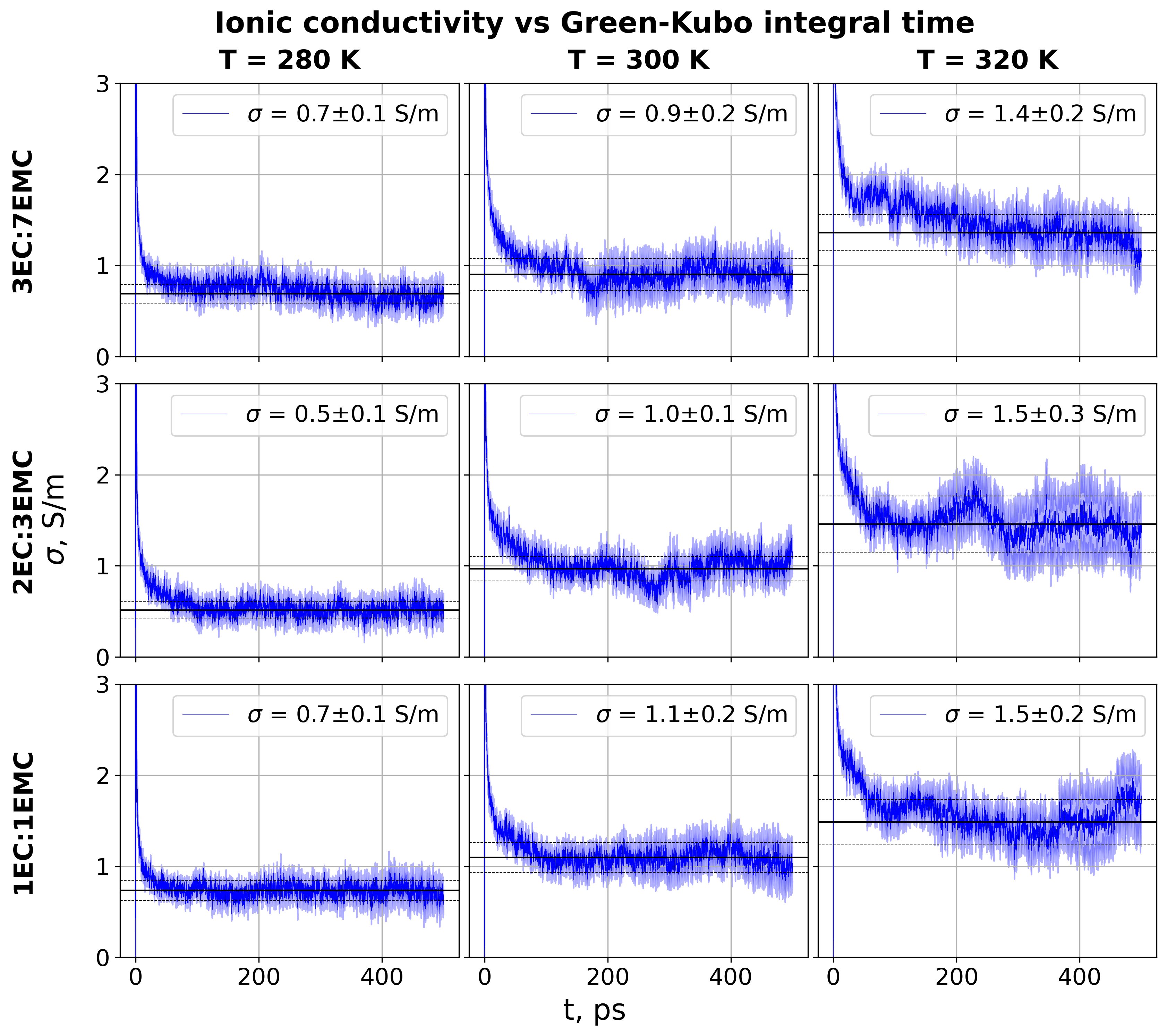}
    \caption{Dependence of ionic conductivity $\sigma$ on correlation time $t$ obtained with MTP, averaged over independent trajectories for each studied system and temperature.}
    \label{fig:composition_temperature_grid}
\end{figure}

\begin{figure}[!ht]
    \centering
    \includegraphics[width=0.8\linewidth]{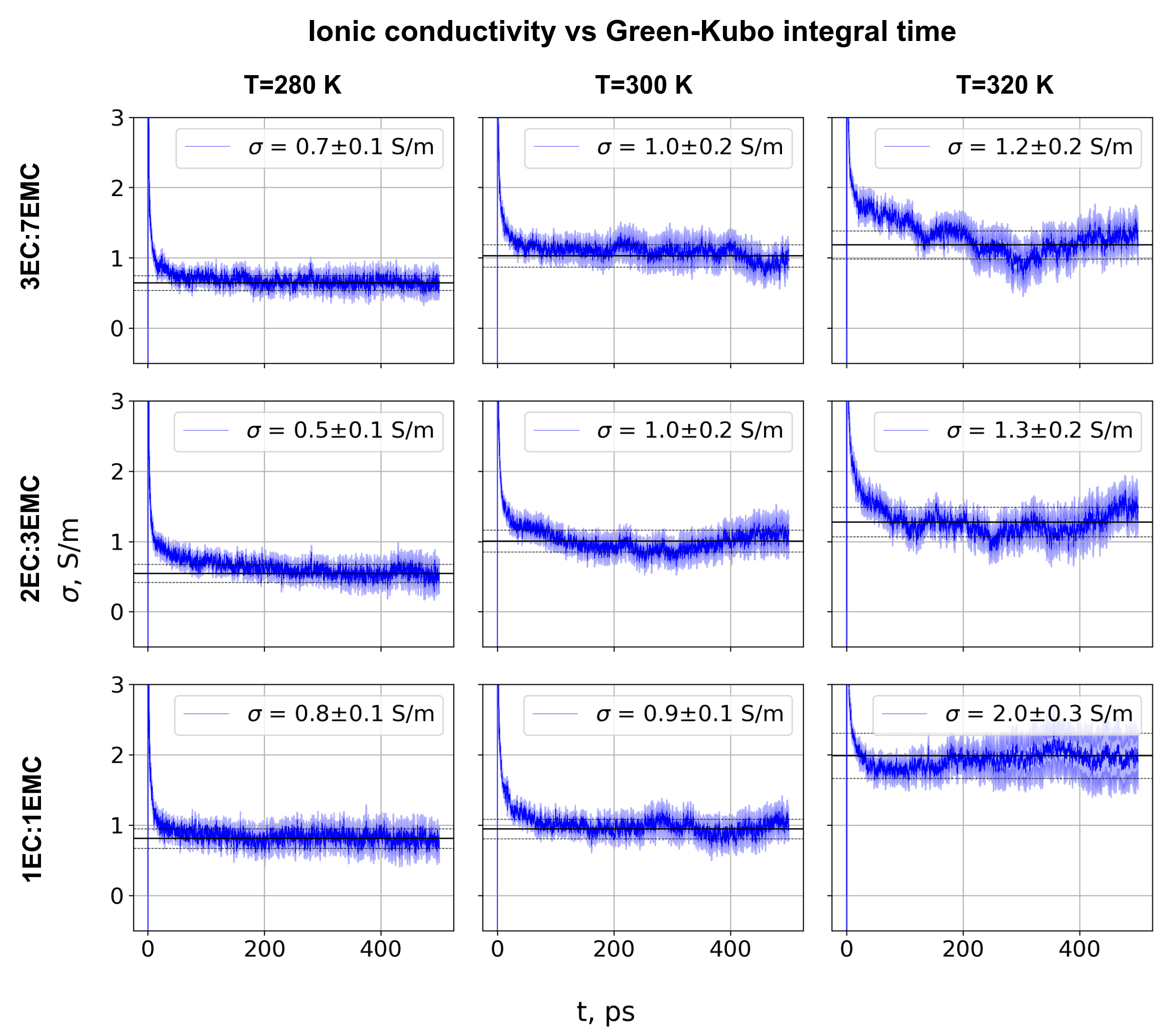}
    \caption{Dependence of ionic conductivity $\sigma$ on correlation time $t$ obtained with MTP-EDQRd, averaged over independent trajectories for each studied system and temperature.}
    \label{fig:SI_ionic_conductivity_vs_corr_time_mtpedqrd}
\end{figure}

Figure \ref{fig:composition_temperature_grid} shows the ionic conductivity versus correlation time for MTP, while Figure \ref{fig:SI_ionic_conductivity_vs_corr_time_mtpedqrd} presents the same for MTP-EDQRd.
The curves, averaged over independent trajectories, are shown for all studied systems and temperatures.

\begin{figure}[!ht]
    \centering
    \includegraphics[width=0.8\linewidth]{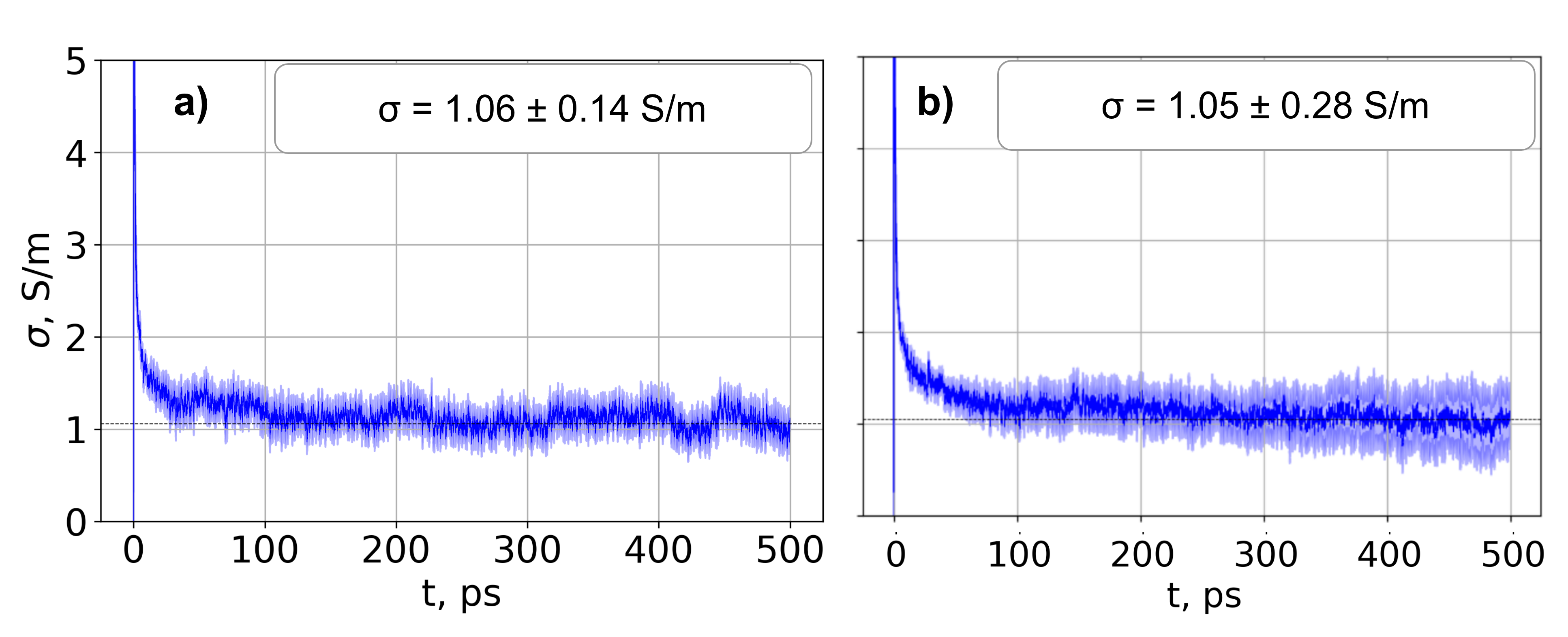}
    \caption{Dependencies of ionic conductivity $\sigma$ on correlation time $t$ obtained with MTP, averaged over independent trajectories for 1EC/1EMC mixture at 300~K a) for cell with 5 LiPF$_6$ ion pairs; b) for cell with 16 LiPF$_6$ ion pairs.}
    \label{fig:SI_finite_size_eff}
\end{figure}

Figure~\ref{fig:SI_finite_size_eff} shows the analogous plot for the 1EC/1EMC mixture at 300~K obtained with the MTP potential for a smaller simulation cell of 19~\AA, containing 5 Li$^+$ and PF$_6^-$ ions (Figure~\ref{fig:SI_finite_size_eff}~(a)) and a bigger simulation cell of 2128 atoms, corresponding to 16 Li$^+$ and PF$_6^-$ ions (Figure~\ref{fig:SI_finite_size_eff}~(a)). In contrast, previous calculations were conducted for an 23~\AA\ cell with 8 Li$^+$ and 8 PF$_6^-$ ions. These results demonstrate that the difference between conductivities obtained for all of the tested cells (Figure \ref{fig:composition_temperature_grid}) is negligible and within the statistical uncertainty across different replicas.

Thus, all conductivity calculations reported in the main text were conducted using 23~\AA\ cell with 8 Li$^+$ and 8 PF$_6^-$ ions. To additionally support these results we also report obtained ionic conductivities in tabular format (Table~\ref{tab:SI_ionic_conductivity_values}).

\begin{table*}[ht]
    \centering
    \caption{Ionic conductivity values obtained with MTP and MTP-EDQRd potentials for simulation cell containing 8 Li$^+$ and 8 PF$_6^-$ ions.}
    \renewcommand{\arraystretch}{1.3}
    \begin{tabular}{|c|c|c|c|c|c|c|c|c|c|}
        \hline
        & \multicolumn{3}{|c}{3EC:7EMC} & \multicolumn{3}{|c|}{2EC:3EMC} & \multicolumn{3}{c|}{1EC:1EMC} \\ \hline
        Temperature, K& 280 & 300 & 320 & 280 & 300 & 320 & 280 &300 &320 \\ \hline
        MTP & 0.7 $\pm$ 0.1 & 0.9 $\pm$ 0.2  & 1.4 $\pm$ 0.2  & 0.5 $\pm$ 0.1  & 0.9 $\pm$ 0.1  & 1.5 $\pm$ 0.3 & 0.7 $\pm$ 0.1 & 1.1 $\pm$ 0.2  & 1.5 $\pm$ 0.3  \\ \hline
        MTP-EDQRd & 0.7 $\pm$ 0.1 & 1.0 $\pm$ 0.2  & 1.2 $\pm$ 0.2  & 0.5 $\pm$ 0.1  & 1.0 $\pm$ 0.2  & 1.3 $\pm$ 0.2 & 0.8 $\pm$ 0.1 & 0.9 $\pm$ 0.1  & 2.0 $\pm$ 0.3  \\ \hline
    \end{tabular}
    \label{tab:SI_ionic_conductivity_values}
\end{table*}

\subsubsection{Diffusion coefficient calculations}

Additionally, we calculated diffusion coefficients for EC:EMC mixtures with ratios of 3:7, 4:6, and 1:1 using the MTP-EDQRd potential. Simulations were performed at 280, 300, and 320~K, with 9--13 replicas per state point. The mean-squared displacement (MSD) curves were averaged across replicas, and the diffusion coefficients were obtained from the three-dimensional Einstein relation:

\begin{equation}
D_i = \frac{1}{6}\frac{\mathrm{d}}{\mathrm{d}t}
      \left\langle \Delta r_i^2(t) \right\rangle .
\end{equation}

The slope was determined by linear fitting over 500-2500 ps of the production time. The phosphorus atom was used to track PF$_6^-$ as its position deviated from PF$_6^-$ ion's center of mass by only 0.04\AA. For the solvent, atoms belonging to EC and EMC were analyzed as one combined group; separate molecular diffusion coefficients for EC and EMC were not calculated. The results are presented in the figure below:

\begin{figure}[!ht]
    \centering
    \includegraphics[width=1\linewidth]{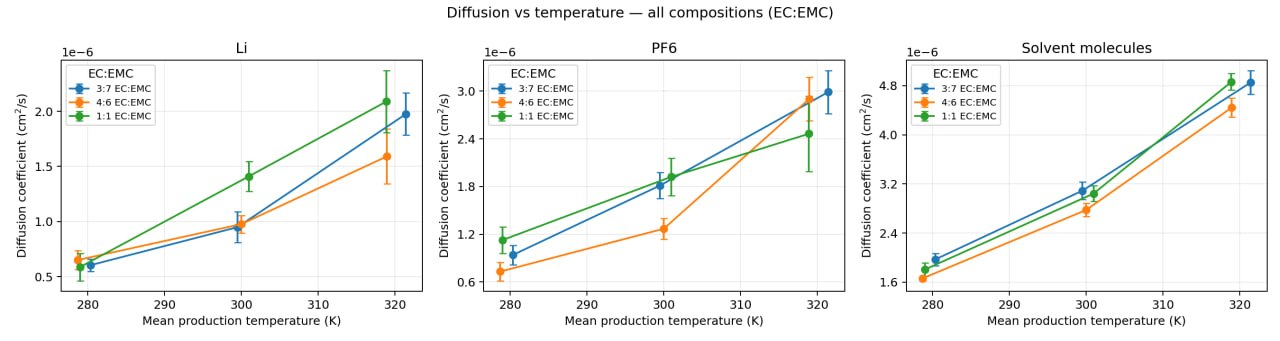}
    \caption{Temperature dependence of the diffusion coefficients of Li$^+$, PF$_6^-$, and the combined EC+EMC solvent-atom group for three EC:EMC compositions, calculated using the MTP-EDQRd potential and the Einstein relation. Symbols show coefficients obtained by fitting the replica-averaged MSD curves. Error bars represent the standard error of diffusion coefficients fitted separately for each replica.}
    \label{fig:SI_diffusion_coefficients}
\end{figure}

\subsubsection{Effect of charge Gaussian smearing on MTP-QRd MD stability}

We also investigated whether introducing Gaussian charge smearing to the fixed-charge QRd model would improve MD stability and enable ionic conductivity calculations. To this end, we applied the same training procedure used for the point-charge MTP-QRd potential (MTP-QRd$_p$) to train a Gaussian-charge MTP-QRd$_g$ model. We then compared the average stable MD simulation times under the same conditions as those used for the ionic conductivity calculations. MTP-QRd$_g$ remained stable for an average of 0.6 ns, compared with 0.7 ns for MTP-QRd$_p$. This result suggests that Gaussian charge smearing alone does not improve the stability of MTP-QRd and that the enhanced stability of MTP-EDQRd cannot be attributed solely to the use of Gaussian charges.

\subsubsection{Fidelity of MLIP ionic conductivity calculations performed outside of the training domain}
To investigate the fidelity of MLIP simulations outside the training domain during ionic conductivity calculations, we evaluated the extrapolation grade for 520 frames sampled from the production trajectories of LiPF$_6$ solutions in 3EC:7EMC, 2EC:3EMC, and 1EC:1EMC at all considered temperatures. Our results indicate that for the 3EC:7EMC solution, configurations predominantly remained within the interpolation regime, as only 4 out of 160 structures exhibited an extrapolation grade higher than the selection threshold used in the AL protocol.

However, for the 2EC:3EMC and 1EC:1EMC compositions, most sampled structures exceeded the extrapolation threshold. Consequently, although the MD simulations remained stable and the calculated ionic conductivities were close to experimental values, these results should be interpreted with caution, as the simulations predominantly sampled configurations outside the interpolation regime.

\subsubsection{Supporting information on structural analysis of the training set and ionic conductivity MD trajectories}

For the classification of ion pairs in the training set and in the MD simulations used to compute the ionic conductivity, we adopted the following procedure. First, the positions of the Li atoms were extracted from the trajectories, and any molecule with at least one atom within 3~\AA\ of a Li atom was assigned to the solvation shell of the corresponding Li$^+$ ion. The resulting solvation shells were then classified as contact ion pairs (CIPs) or solvent-separated ion pairs (SSIPs) according to the presence or absence of a PF$_6^-$ ion in the shell, respectively. Finally, aggregate configurations were identified as CIPs sharing at least one common PF$_6^-$ ion, while networks were identified as at least two ion pairs sharing a common solvent molecule---with at least one of its oxygen atoms lying within the solvation shells of two Li$^+$ ions---but without common PF$_6^-$ ion containing these pair of Li$^+$ atoms.

\begin{figure}[!ht]
    \centering
    \includegraphics[width=1\linewidth]{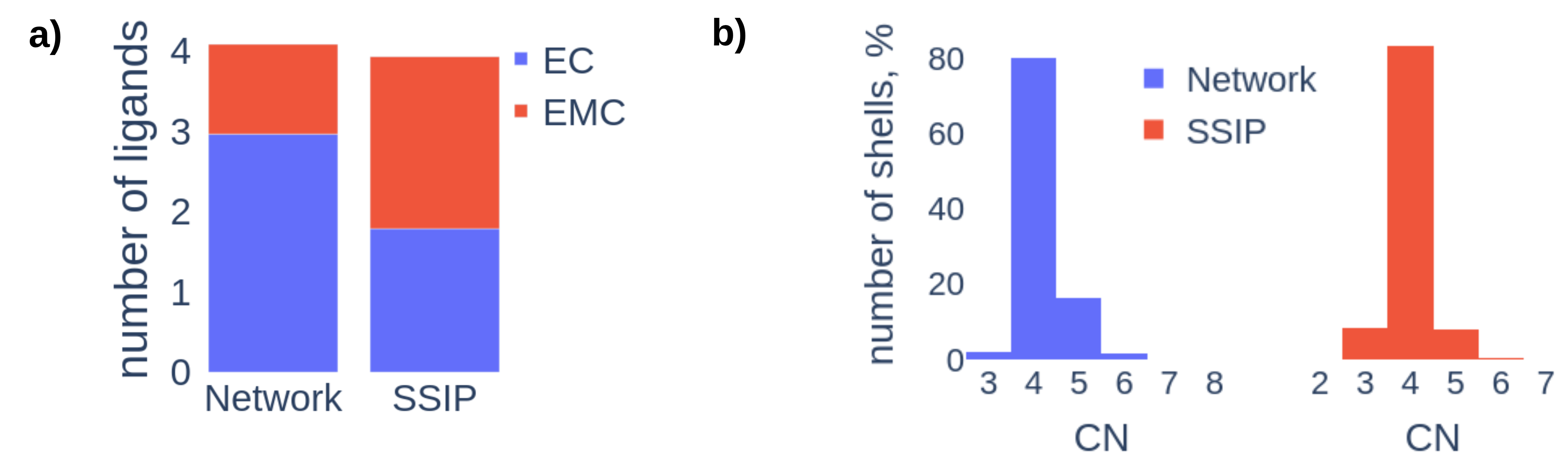}
    \caption{
    a) Differences in solvation shell composition between SSIPs and SSIPs forming networks (Networks) from MTP trajectories; 
    b) corresponding differences in coordination number (CN) distributions.}
    \label{fig:long_trj_analysis_SI}
\end{figure}

Figure \ref{fig:long_trj_analysis_SI} presents additional analysis of MD trajectories underlying ionic conductivity calculations with MTP. Panel (a) shows differences in solvation shell composition between solvent separated ion pairs (SSIPs) and SSIPs forming networks (denoted as Networks), while panel (b) shows corresponding differences in coordination numbers (CNs) distribution, indicating higher Li$_{^+}$ coordination in Networks.

\printbibliography

\end{document}